\documentclass[traditabstract,longauth]{aa}  

\usepackage{graphicx}
\usepackage{txfonts}
\usepackage{float}
\usepackage{xcolor} 

\usepackage{natbib}
\usepackage[switch]{lineno}
\usepackage{graphicx}
\usepackage{subcaption}
\usepackage{hyperref}
\hypersetup{
    colorlinks=true,
    citecolor={blue}
}

\begin{document} 
   \title{Self lensing in BH-NS binaries as progenitors for periodically repeating Fast Radio Bursts}
   \titlerunning{Periodically repeating FRBs: BH-NS self-lensing scenario}  

   \author{L.E. Rodr\'iguez \fnmsep\thanks{\email{lsrodrig@uc.cl}},
          \inst{1}
          M. Cruces, 
          \inst{2}
           G. Voisin 
          \inst{3}, 
          J. Vera-Casanova 
          \inst{2},
          C.A. Braga 
          \inst{1}
          }

  \authorrunning{Rodríguez et al.}         

   \institute{Instituto de Astrof\'{i}sica, Pontificia Universidad Cat\'{o}lica de Chile, Av. Vicu\~{n}a Mackenna 4860, Macul, Santiago, Chile
        \and
         Centre of Astro-Engineering, Pontificia Universidad Católica de Chile, Av. Vicuña Mackenna 4860, Santiago, Chile
         \and
   LUX, Observatoire de Paris, Université PSL, Sorbonne Université, CNRS, 92190 Meudon, France
             }

   \date{Received XXXX XX, XXXX; accepted XXXX XX, XXXX}
 
  \abstract
   {Fast radio bursts (FRBs) are bright extragalactic transients whose origin remains uncertain. The discovery of repeating FRBs --some with {cyclic} activity windows-- {open new prospects for testing their progenitors}. Neutron stars (NSs) are strong candidates, as they naturally emit radio pulses with temporal and spectral properties resembling those of FRBs. This interpretation was reinforced by the  detection of an FRB-like burst from the Galactic magnetar SGR 1935+2154. Nevertheless, the pulse energy of typical radio-emitting NSs remains several orders of magnitude below that of even the faintest observed FRBs.}
   { {We explore {the scenario of} a binary system in which a NS orbits a black hole (BH).} In this scenario, the NS’s radio pulses are gravitationally amplified through self-lensing, {enhancing the observed energy and producing characteristic signatures in the energy distribution, activity windows, and waiting times that can be contrasted with observations } }
   {{We implement a simple NS--BH binary model with a circular orbit undergoing Lense-Thirring precession}, where the BH’s spin drags the surrounding spacetime and drives orbital nodal precession.
   We then compute how the radio pulses are gravitationally lensed along the orbit, incorporating effects such as pulse splitting and time delays.}
   {Pure Keplerian, non-precessing orbits produce activity windows several orders of magnitude shorter than those observed in periodic FRBs. More promising are post-Keplerian orbits subject to Lense–Thirring precession, which can lead to activity windows ranging from a few to tens of days. {We show that lensing by the BH transforms an intrinsic log-normal NS pulse-energy distribution, producing a high-energy power-law tail extending over up to four orders of magnitude}. Interpreting the observed millisecond waiting times between consecutive FRB bursts as lensing-induced image delays, {we infer BH masses of $10^{3}$--$10^{4}$ M$_{\odot}$, placing them in the intermediate-mass (IMBH) range}.} 
   {{Although this is a promising scenario, the amplification obtained in our model is insufficient to fully bridge the energy gap between typical Galactic magnetar radio pulses and the bursts observed from FRBs. In addition, the non-detection to date of NS-IMBH systems, and hence their highly uncertain and likely low expected abundance, suggests that this channel is rare.}}

   \keywords{gravitational lensing: weak –stars: magnetars – stars: neutron – radio continuum: general
            }

\maketitle
%

\section{Introduction}

Fast radio bursts (FRBs) are radio transient extragalactic events with timescales as short as a fraction of a millisecond. To date, they have been detected only at radio frequencies up to 8 GHz \citep{Gajjar2018, Zhang2020}.  The exact origins of FRBs remain unknown; however, their high isotropic equivalent luminosity ($\sim10^{38}-10^{46}$ erg s$^{-1}$), timescales ($\sim10^{-6}-10^{-3}$ s) and pulse morphology suggest an association with high-energy events involving compact objects, with neutron stars (NSs) being one of the most promising progenitors.

The discovery of repeating Fast Radio Bursts (rFRBs) represented a pivotal breakthrough in the field because they exclude cataclysmic progenitor scenarios that would destroy the source as the origin for all FRBs \citep{Spitler2016}. Unlike their non-repeating counterparts, rFRBs enable studies of their energy distributions, waiting times and repetition patterns \citep{cruces,Li2021,Braga2025}.

A subset of rFRBs shows periodic activity windows (hereon referred to as periodic rFRBs or prFRBs, indistinctively). Currently, only three such periodic sources have been reported: \object{FRB 20121102A} with a period $P \sim 160$ days \citep{Rajwade,cruces,Braga2025}, \object{FRB 20180916B} with $P \sim 16$ days \citep{CHIME2020periodic}, and \object{FRB 20240209A} with $P \sim 126$ days \citep{Pal2025}. For these three sources, burst activity is temporally constrained to windows covering $\sim$ 50\%, $\sim$ 30\%, and $\sim$ 40\% of their cycles for \object{FRB 20121102A}, \object{FRB 20180916B}, and \object{FRB 20240209A}, respectively.

Among known prFRBs, \object{FRB 20121102A} has the longest time span, as it is the first repeater discovered  \citep{Spitler}. Analysis of its bursts reveals a bimodal energy distribution \citep{Li2021}. At lower energies ($E < 10^{38}$~erg), the distribution follows a lognormal function centred at $\sim 4.8 \times 10^{37}$~erg, while the high-energy component is well-fitted by a generalised Cauchy function that approaches a power law with index $\gamma \approx -1.37$ \citep{Li2021}. On the other hand, the waiting time distribution between consecutive bursts reveals two lognormal-like structures \citep{Li2021}, with primary and secondary peaks at $\sim 70$~s and $\sim 3.4$~ms, respectively. This two-lognormal set of waiting times appears consistent across other repeating FRBs  \citep{Li2021,Xu2022,Zhang2023a}. Throughout the activity window, the burst rates display a clear trend, peaking near the center of the activity window and exhibiting a roughly Gaussian profile around this maximum \citep{EspinozaDupouy2025}.

The analysis of the features of some FRBs has been interpreted as evidence that their sources could be embedded in complex magneto-ionic environments \citep[e.g.][]{AnnaThomas2023}. Information about these environments can be obtained from the interaction of the radio signal with plasma along its path, which introduces delays in the propagation velocity and a rotation of the polarization angle in the presence of a magnetic field aligned with the line of sight (LoS). These effects are quantified by the dispersion measure (DM) 
and the rotation measure  (RM)
, respectively. 
After accounting for the contributions from the host galaxy and the intergalactic medium along the LoS, a significant excess in DM remains in some events, suggesting the presence of plasma in the vicinity of the source. Similarly, rapid variations in RM in some FRBs have been associated with the radio signal propagating through turbulent, magnetized plasma. For certain rFRBs, the observed values of DM and RM are comparable to those measured for the Galactic center magnetar \citep{Desvignes2018}.

In April 2020, the magnetar \object{SGR 1935+2154} entered an active phase, emitting two high-energy radio bursts \citep{Andersen2020}--now classified as \object{FRB 20200428}. These events were roughly two orders of magnitude less energetic than typical FRBs, while still being up to a million times more intense than standard {magnetar} radio pulses. This observational link strongly supports magnetars as the most plausible FRB progenitors. Magnetars are a rare sub-class of NSs (<1\%),  distinguished by their young age ($\lesssim 10^4$\,yr), slow spin periods ($P \gtrsim 1\mbox{--}12$\,s), and the strongest known magnetic fields ($\sim 10^{14}\mbox{--}10^{15}$\,G). Of the $\sim$29 galactic magnetars identified, only 7 have shown radio emission \citep{Rea2025}, making \object{SGR 1935+2154}'s FRB-like bursts particularly significant.

The quest to explain FRBs has generated a wide array of theoretical models, each attempting to address different aspects of the puzzle (see \citealt{Zhang2023} for a review). However, no {model} yet offers a unified explanation for the full range of prFRB phenomena. Some models provide important insights into specific observed features, such as the generation of coherent radio emission or the repetition patterns in prFRBs.

A relevant advance in understanding the source mechanism for the production of FRBs comes from the  work done in \citet{Voisin}. It shows how FRBs could arise from radiation emitted from relativistic plasma moving along open magnetic field lines of a NS. This mechanism naturally reproduces the observed FRB morphology, especially in magnetar configurations with substantial toroidal magnetic-field components. This framework provides a solid quantitative link between magnetar physics and geometry with FRB emission properties. 

{Beyond the question of how FRBs are generated, there is an ongoing debate regarding the conditions required for the resulting coherent radio waves to escape a magnetar's magnetosphere without being absorbed or strongly modified by the surrounding plasma. \citet{Beloborodov2021} argued that radio waves emitted within the closed-field-line region of a quiescent magnetar face severe non-linear absorption as they propagate through the dense magnetospheric plasma, casting doubt on emission scenarios confined to that region. Subsequent work has shown that this constraint can be relaxed for emission generated on open field lines or in the presence of a relativistically outflowing plasma \citep{Lyutikov2024}, partially reconciling magnetospheric models with the observed FRB luminosities. Importantly, \citet{Xiao2024} showed that propagation through the magnetosphere can preferentially suppress bursts separated by short time lags.}

The most accepted models to explain the cyclic behavior of prFRBs point to two leading scenarios: precession of a magnetar's spin axis, and a binary system where a companion object intervenes, modulating the NS's emission.

{The precession models suggest that a highly energetic magnetar emits a beam directed toward Earth, with its orientation changing over time due to spin-axis precession, thereby producing the observed activity windows. This possibility was proposed shortly after the discovery of periodic activity in \object{FRB~20180916B} \citep{CHIME2020periodic}, with models invoking free precession of a magnetically deformed neutron star \citep{Levin2020,Zanazzi2020}. More recent works have further explored how the evolution of the internal magnetic field, spin period, and obliquity can affect the precession dynamics and the resulting activity windows \citep[e.g.][]{Feng2024}. Most of these models omit the presence of superfluid vortices in the NS’s interior. According to established NS physics, these vortices tend to dampen precessional motion, acting like a gyroscope \citep{Akgun2006,Shaham1977}, making it difficult to systematically explain the wide range of periodicities observed in FRBs. 

{In particular, the paper reporting the discovery of periodic activity in \object{FRB~20180916B} \citep{CHIME2020periodic} noted that a stellar-mass or intermediate-mass BH companion could produce periodically observed bursts through relativistic spin--orbit coupling.}

{Binary models involving NSs with massive companions—stellar companions or BHs with accretion disks—propose that FRBs arise from the NS's radio beam propagating through optically thin, spiral funnel-shaped regions carved into the companion's wind; activity windows occur when these funnels align with the observer's LoS, while FRBs are obscured when they rotate away \citep{Ioka,Lyutikov,Wada2021}. However, these models predict periodic DM/RM modulation not yet conclusively observed in prFRBs, and they do not address the NS--FRB energy gap, since the wind only modulates the optical depth passively without contributing to the emission mechanism itself.}

To solve the energy gap problem, the gravitational lensing by compact objects has recently been considered as a viable alternative. \cite{DallOsso2024} proposed that FRB-like signals, originating from co-rotating hot spots anchored in NS magnetospheres, can be significantly amplified by the strong gravitational field of the NS itself. Given the strong dependence of lensing on viewing angle, they argue that prFRBs could be explained by extremely rare and finely tuned observer orientations, resulting in high probabilities of amplification and frequent burst detection.

{Observationally, FRBs have also been proposed as sensitive probes of compact gravitational lenses because their short durations allow small lensing-induced time delays to be resolved. Recently, \citet{Xiong2026} reported evidence for a possible gravitationally lensed event in \object{FRB~20190320B}, with the inferred lens mass of order a few hundred solar masses. Although this interpretation requires further confirmation, it illustrates that image splitting by compact lenses can leave observable temporal signatures in FRB data.}

{A related lensing phenomenology can occur in binary systems, where radiation emitted by one component is amplified by the gravitational field of its compact companion. This binary self-lensing configuration was discussed by \citet{Maeder1973}. In nearly aligned source--lens--observer configurations, the compact companion can amplify the emitted signal by factors of order $\sim10^4$ or higher.}

We explore the scenario in which prFRBs originate from the self-lensing of radio pulses emitted by a NS through its black hole (BH) companion. This configuration naturally produces episodes in which the radio beams from the NS are re-collimated toward Earth when the NS is eclipsed by the BH. 
As a result, one would expect to observe multiple flashes following a repetitive pattern determined by the orbital geometry of the NS around the BH.

Sect.~\ref{sec:Methodology} describes the methodology and presents the set of equations that compose the model employed in this work. Sect.~\ref{sec:Results} discusses the results of our simulations for orbits both without and with nodal precession, comparing them with the observational data. Finally, Sect.~\ref{sec:conclusions} summarizes our main conclusions.

\section{Methodology}\label{sec:Methodology}

We develop a simplified model consisting of a NS orbiting a rotating BH. In this system, the NS periodically emits radio bursts that experience signal amplification through gravitational self-lensing as they propagate near the BH. Our objective is to determine whether a pulse from a radio emitting NS could be sufficiently amplified to be observed as {an} FRB and to identify the orbital phases during which this could occur. 

We assume that the radio pulse energy distribution from neutron stars follows a log-normal distribution, consistent with what is observed in pulsars \citep{Burke2012}. We acknowledge that in the case of magnetars — the most promising candidates — the emission of SGR 1935+2154, the FRB link, differs significantly from the regular pulsar radio emission. These bursts are not periodically emitted and exhibit energies several orders of magnitude higher, suggesting distinct trigger mechanisms. {Nevertheless, we adopt the distribution of regular pulsar pulses as a working model for magnetar-like radio emission, in order to map the amplification zones and estimate upper limits on burst energies and occurrence rates.}

{As the bursts propagate near the black hole, their profiles can be modified by gravitational lensing. When the neutron star becomes nearly aligned with the black hole and the observer, a burst may be split into two lensed images: a primary component and a secondary {fainter} echo. The arrival-time offset between these components is a lensing-induced image delay, which includes both the geometrical path-length difference and the gravitational time delay associated with the BH {mass} \citep{Maeder1973,schneider}. This effect can increase the effective burst detection rate, but only when the amplified energy of the secondary component also exceeds our detection threshold.}

The amplification factor is calculated following the self-lensing framework of \citet{Maeder1973}, which explicitly accounts for the finite size of the source {and is based on the Schwarzschild metric}. {We adopt this formalism as an order-of-magnitude approximation that captures the leading geometrical dependence of finite-source self-lensing. The model operates close to the transition toward the strong-lensing regime, where deviations from the weak-deflection treatment may become important \citep[e.g.][]{Bakala2023}. For our fiducial configurations discussed below, the orbital separation is typically $a/r_s\sim 200$, where $r_s=2GM_{\rm BH}/c^2$ {is the Schwarzschild radius of the BH}, corresponding to an Einstein ring scale $R_E\simeq \sqrt{2r_sa}\sim 20r_s$. Therefore, the weak-deflection approximation should be regarded as an order-of-magnitude description. We note that \cite{Maeder1973} assumes isotropic emission, whereas radio emission from NSs is typically highly collimated. {In this sense, our isotropic treatment should be regarded as a conservative reference normalization for the emitted energy. Directional effects associated with beaming, spin-axis precession, and the detailed magnetospheric emission geometry are left for future work.} 

{For our baseline analysis, we establish {a reference amplification} threshold requiring an amplification of at least $\sim100\times$, roughly corresponding to the energy gap between the brightest FRB-like radio bursts detected from the Galactic magnetar SGR~1935+2154 \citep{Andersen2020} and the faintest extragalactic FRBs. From the subset of simulated pulses that meet this threshold, we construct energy and waiting-time distributions for comparison with prFRB observations.}

{The model includes as free parameters the BH mass and spin, the NS cadence period, the orbital period, inclination, and eccentricity. In the fiducial simulations presented below, we set $e=0$.}

\subsection{Model}

To characterize the burst amplification mechanisms in the NS-BH binary system, we formulate a simple model that simultaneously addresses orbital dynamics and gravitational lensing phenomena. The framework combines a post-Newtonian treatment of the NS's orbit, incorporating both Lense-Thirring frame-dragging and precession of periapsis; the amplification and splitting of radio pulses during close alignments through gravitational self-lensing; and  relativistic time-delay effects imprinted on the observed signals. This comprehensive approach enables us to systematically connect the binary's orbital kinematics and geometry with the observed burst properties, where the NS's motion, position, and radio emission govern both the temporal structure of bursts and their flux modulation.

\subsubsection{Orbit geometry}\label{subsec:equations}

We begin by outlining the construction of the orbital model, incorporating the relativistic effects that drive deviations from a purely Keplerian orbit; the Keplerian case is recovered by setting all precession terms to zero.

The dominant effect is the frame dragging produced by the BH's rotation, which drives the Lense--Thirring precession (LTp) of the longitude of the ascending node, $\Omega$. Following \citet{Merritt2013}, the average precession rate is given by
\begin{equation}\label{LTpres}
\biggl \langle  \frac{d\Omega}{dt} \biggr \rangle= \frac{2G^2M_{\mathrm{BH}}^2}{c^3a^3(1-e^2)^{3/2} }\chi, 
\end{equation}
where $M_{\mathrm{BH}}$ is the BH mass, $\chi$ its dimensionless spin parameter, and $a,e$ the semi-major axis and eccentricity of the orbit. 

 {We adopt $\chi=0.9$ as a fiducial high-spin case to illustrate the effects of LTp. Such a high spin is not required for the model to reproduce the observed activity cycle. The impact of lower spins is discussed in Sect.~\ref{subsec:LTOrbits}, while Table~\ref{tab:spin_dependence} presents a parameter combinations spanning $0.1 \leq \chi \leq 0.9$ that preserve the adopted cycle and the activity fraction.}

LTp causes the ascending and descending nodes to precess and periodically intersect the caustic line, on a characteristic timescale
\begin{equation}\label{T_LT}
\tau_{\mathrm{LT}}= 2\pi \biggl \langle  \frac{d\Omega}{dt} \biggr \rangle^{-1}= \frac{\pi(1-e^2)^{3/2}  c^3a^3}{G^2M_{\mathrm{BH}}^2\chi}.
\end{equation}
{For a circular, prograde equatorial orbit in the Kerr metric, the orbital period measured in coordinate time is \citep{Bardeen1972}
\begin{equation}\label{T_ORB_1}
\tau_{\rm Orb}
= 2\pi\sqrt{\frac{a^3}{GM_{\rm BH}}}
+ \frac{2\pi\chi GM_{\rm BH}}{c^3},
\end{equation}
where we identify the Boyer--Lindquist orbital radius with $a$ in our nearly equatorial approximation. For our fiducial parameters, the spin-dependent correction is approximately $10^{-4}$ of the Keplerian term. We therefore adopt the leading-order Keplerian expression, which we also use in the derivations below. Since this orbital period is much shorter than $\tau_{\rm LT}$, the fast orbital motion occurs within the slowly varying LTp cycle, and each node crossing of the caustic line generates one activity episode.}

{We therefore associate the full FRB activity cycle with half the LT timescale,
\begin{equation}\label{T_FRB_LT}
\tau_{\mathrm{FRB}}=\frac{1}{2}\tau_{\mathrm{LT}}.
\end{equation}}
For a given BH mass $M_{\rm BH}$ and spin $\chi$, Eq.~(\ref{T_LT}) then determines the semi-major axis $a$; combining this with Eq.~(\ref{T_ORB_1}) yields the orbital period
\begin{equation}\label{TORB}
{\tau}_{\mathrm{Orb}} = \sqrt{ \frac{8 \pi G M_{\mathrm{BH}} \chi \, \tau_{\mathrm{FRB}}}{c^3 (1 - e^2)^{3/2}} }.
\end{equation}
Meanwhile, the ratio between the active and inactive phases of the cycle depends on the orbital inclination $i$ and $M_{\rm BH}$ (Sect.~\ref{subsec:LTOrbits}).

In addition to LTp, the orbit is subject to secular precession of the periapsis $\omega$, combining the Schwarzschild contribution and the leading spin--orbit (Lense--Thirring) contribution associated with the BH angular momentum \citep[e.g.][]{Merritt2013}. The full orbit is then built by combining the underlying Keplerian ellipse with the periapsis precession $\omega(t)$, the nodal precession $\Omega(t)$ given by Eq.~(\ref{T_LT}), and the orbital inclination $i$ with respect to the BH spin axis, finally projected into the observer-frame coordinates $x_{\rm Obs}, y_{\rm Obs}, z_{\rm Obs}$. The explicit Keplerian radial equation, the periapsis-precession rate, and the full sequence of coordinate transformations used to construct $x_{\rm Obs}, y_{\rm Obs}, z_{\rm Obs}$ are given in Appendix~\ref{app:orbit}. The parameters $\Omega, \omega, \nu, i$ of the orbit can be visualized in Fig.~\ref{fig:coordinatesystem}. Throughout this work, we define the scale of the system as the ratio between the semi-major axis of the orbit and the Schwarzschild radius of the BH, $a/r_s$.

The lifespan of the system is computed as an order of magnitude from the merger time by gravitational wave emission, e.g. \citet{Merritt2013}
\begin{equation}\label{lifespan}
   \tau_{\mathrm{GW}}= \frac{5}{256} \frac{c^5}{G^3} \frac{a^4}{m_{\mathrm{NS}}M_{\mathrm{BH}}(m_{\mathrm{NS}}+M_{\mathrm{BH}} )},
\end{equation}
where $M_{\mathrm{BH}}$ is the mass of the BH and $m_{\mathrm{NS}}$ is the mass of the NS, and $a$ is the semi-major axis.
For the range of masses and orbital separations of interest—intermediate-mass BHs and NSs orbiting at hundreds of {Schwarzschild} radii—we obtain $\tau_{\mathrm{GW}} \sim \mathcal{O}(10^3 ~\rm yr)$.
This timescale is reasonable for the ages of FRBs, as FRB 20121102A has been detected for at least a decade, setting a firm lower limit.

\subsubsection{Amplification}\label{subsec:amplification}

Once the orbit is defined, it is possible to compute the amplification factor $\mathcal{A}$ that depends on the alignment of the NS with the BH. {The closer the NS is to the caustic line, the larger the magnification of the signal}; see Fig. \ref{fig:coordinatesystem}.

The gravitational deflection angle {$\alpha = 2r_s/r_{\gamma}$}, where {$r_{\gamma}$} is the minimal distance of the geodesic of the photon to the BH, determines the light bending near the BH. This relativistic deflection focuses the radiation toward the observer, producing a magnified image of the NS; see Fig. \ref{fig:magni}. 

We quantify the signal amplification $\mathcal{A}$ through the solid angle ratio:
\begin{equation}\label{eq:ampli}
\mathcal{A} \equiv \frac{\Omega_{\text{lensed}}}{\Omega_{\text{source}}} \approx \left(\frac{R_{\text{lensed}}}{R_{\text{source}}}\right)^2,
\end{equation}
where $\Omega$ represents the respective subtended solid angles and $R$ the characteristic sizes of the source and the lensed image. 
To compute the lensed area, it is necessary to integrate all light rays emitted from the NS’s surface that are deflected by an angle $\alpha$ such that they can reach Earth. This deflection angle depends on the relative position of the NS with respect to the BH. 

{In the literature, the amplification factor $\mathcal{A}$ is typically expressed as a function of the projected source--lens separation $b$ \citep{Maeder1973,Liebes1964}}. Here, $b$ is the distance between the projected position of the NS and the caustic line, measured in the plane perpendicular to the line of sight. In our observer-frame coordinates, this distance is $b=\sqrt{x_{\rm Obs}^2+z_{\rm Obs}^2}$; see Fig.~\ref{fig:AmplificationModel2}. This quantity should be distinguished from $r_\gamma$, which denotes the closest approach of an individual photon trajectory to the BH.

{Using this projected separation, we compute the amplification with a piecewise Maeder--Liebes prescription. The finite-source { Maeder \citep{Maeder1973} } expression is retained close to the caustic and around the contact region, while the {Liebes \citep{Liebes1964}} approximation is used only in the far outside-contact regime, where it accurately reproduces the finite-source solution. We therefore use}

{
\begin{equation}
\mathcal{A}(b)=
\begin{cases}
\dfrac{2}{\pi}\sqrt{1+\frac{8r_sa}{R^2}} E\left(\dfrac{b}{R}\right), & b<R,\\
\frac{2}{\pi}
\left(
1+\frac{2r_s a}{R^2}
\right)
\sin^{-1}
\left[
\frac{1}{\sqrt{1+2r_s a/R^2}}
\right]
+
\frac{2}{\pi}
\sqrt{\frac{2r_s a}{R^2}}, & b=R,\\
\dfrac{2}{\pi}\sqrt{1+\frac{8r_sa}{R^2}} \dfrac{b}{R}
\left[
E\left(\dfrac{R}{b}\right)-
\left(1-\dfrac{R^2}{b^2}\right)
K\left(\dfrac{R}{b}\right)
\right], & R<b<3R,\\
\mathcal{A}_{\rm L}(b), & b\geq 3R.
\end{cases}
\label{eq:A_piecewise}
\end{equation}}
{where $E$ and $K$ are complete elliptic integrals of the second and first kind, respectively. The first branch corresponds to the near-caustic finite-source Maeder amplification. In the limit of exact alignment, $b=0$, the finite-source amplification saturates at}
{\begin{equation}\label{A_max}
\mathcal{A}_{\max} \approx \sqrt{\frac{8r_sa}{R^2}}.
\end{equation}}
{where in the isotropic case we assume $R=R_{\rm NS}$. The second branch in Eq.~(\ref{eq:A_piecewise}) is the contact value at $b=R$, which reduces to $\mathcal{A}_{\rm c}\simeq(2/\pi)\mathcal{A}_{\max}$ in the high-amplification limit. The third branch is the outside-contact Maeder expression, while $\mathcal{A}_{\rm L}$ is the Liebes approximation,}
\begin{equation}\label{A_liebes}
\mathcal{A}_{\rm L}(b)=
\frac{1}{2}
\left[
\sqrt{1+\frac{8r_sa}{b^2}}
+
\left(1+\frac{8r_sa}{b^2}\right)^{-1/2}
\right],
\end{equation}
{which is valid for $b\gtrsim3R$. This prescription avoids applying the point-source Liebes approximation in the caustic regime, where finite-source effects regulate the maximum amplification.} Finally, the luminosity $L_{\mathrm{NS}}$ of those pulses is amplified as
\begin{equation}\label{EAmp}
{L_{\rm{Lensed}}= g^4\mathcal{A} L_{\mathrm{NS}},}
\end{equation}

{In Eq.~(\ref{EAmp}) we set the redshift factor to $g\simeq1$ and neglect spin-dependent corrections to the photon deflection, {see Appendix~\ref{app:relativistic_corrections}}. For the fiducial configurations considered here, these effects introduce only small corrections compared with the several orders of magnitude relevant to the lensing amplification.}

\begin{figure}[ht]
    \centering
    
    \begin{subfigure}[b]{\linewidth}
        \centering
        \includegraphics[width=\linewidth]{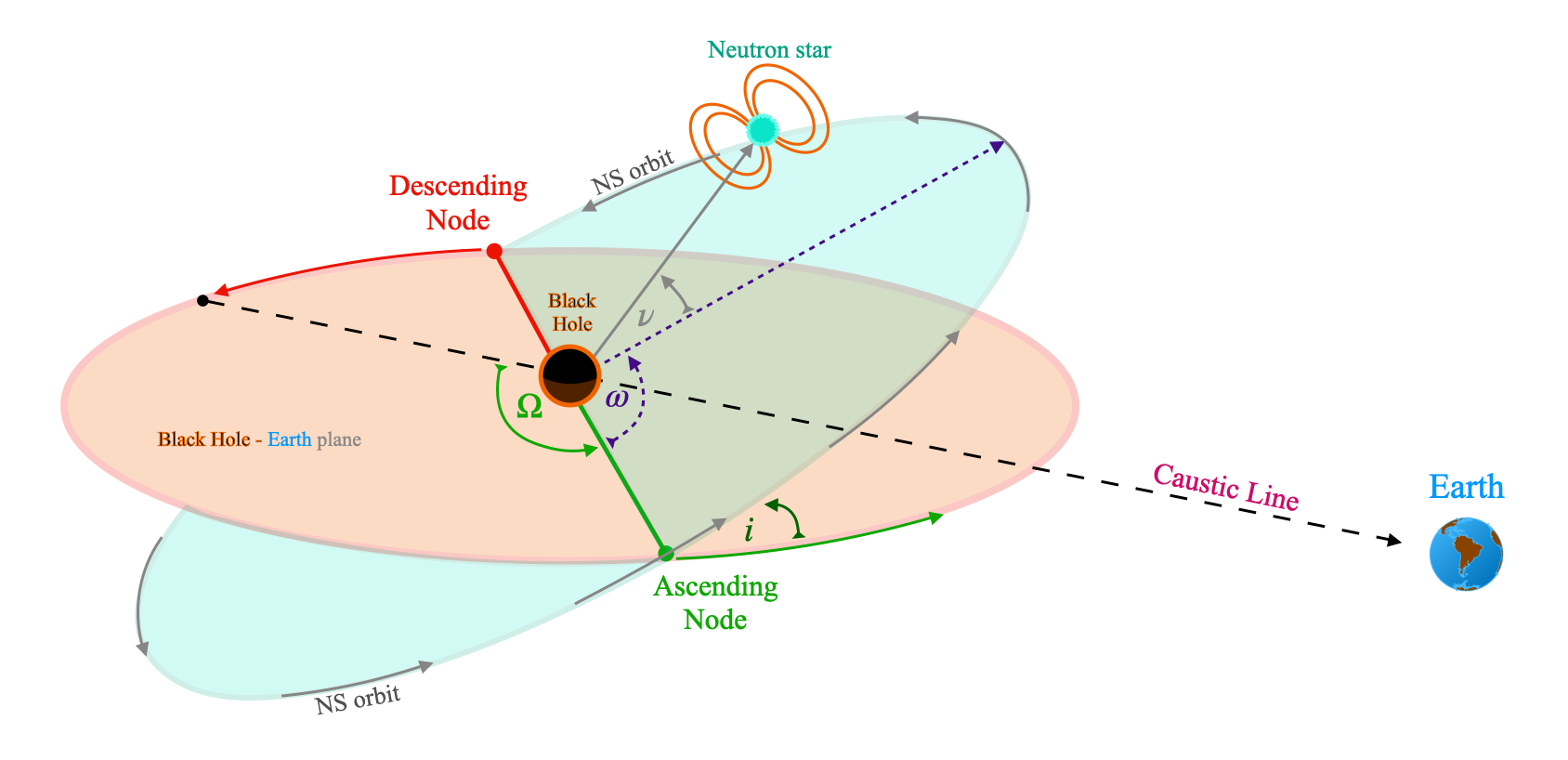}
        \caption{\small{Elliptical orbit of a NS around a BH, with inclination $i$, node precession $\Omega$, periapsis precession $\omega$, and true anomaly $\nu$.}}
        \label{fig:coordinatesystem}
    \end{subfigure}
    \vspace{0.3cm}
    
    \begin{subfigure}[b]{\linewidth}
        \centering
        \includegraphics[width=\linewidth]{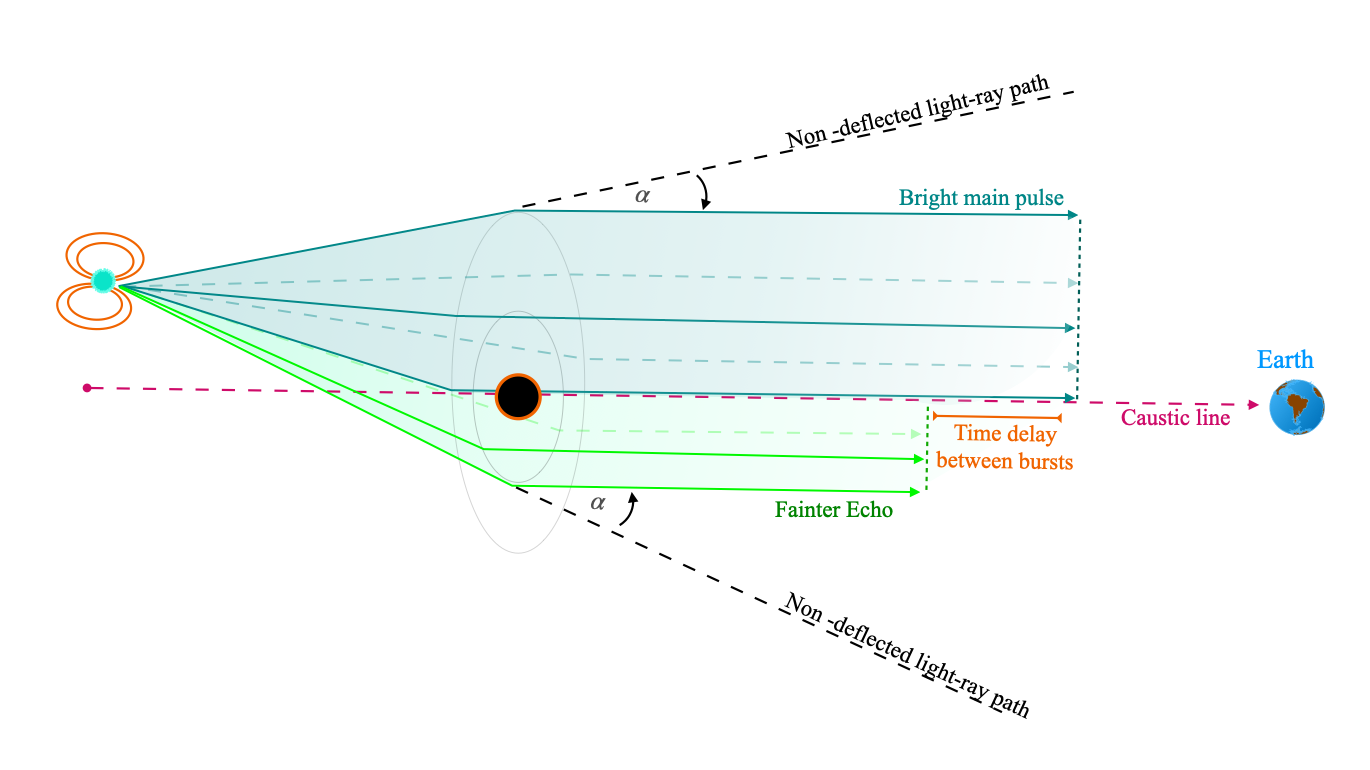}
        \caption{\small{ {Schematic of gravitational lensing: light rays passing near the BH are deflected by an angle $\alpha\simeq2r_s/r_{\gamma}$, producing a primary burst and a delayed secondary echo.}}}
        \label{fig:magni}
    \end{subfigure}
       \vspace{0.3cm}

     \begin{subfigure}[b]{\linewidth}
        \centering
        \includegraphics[width=\linewidth]{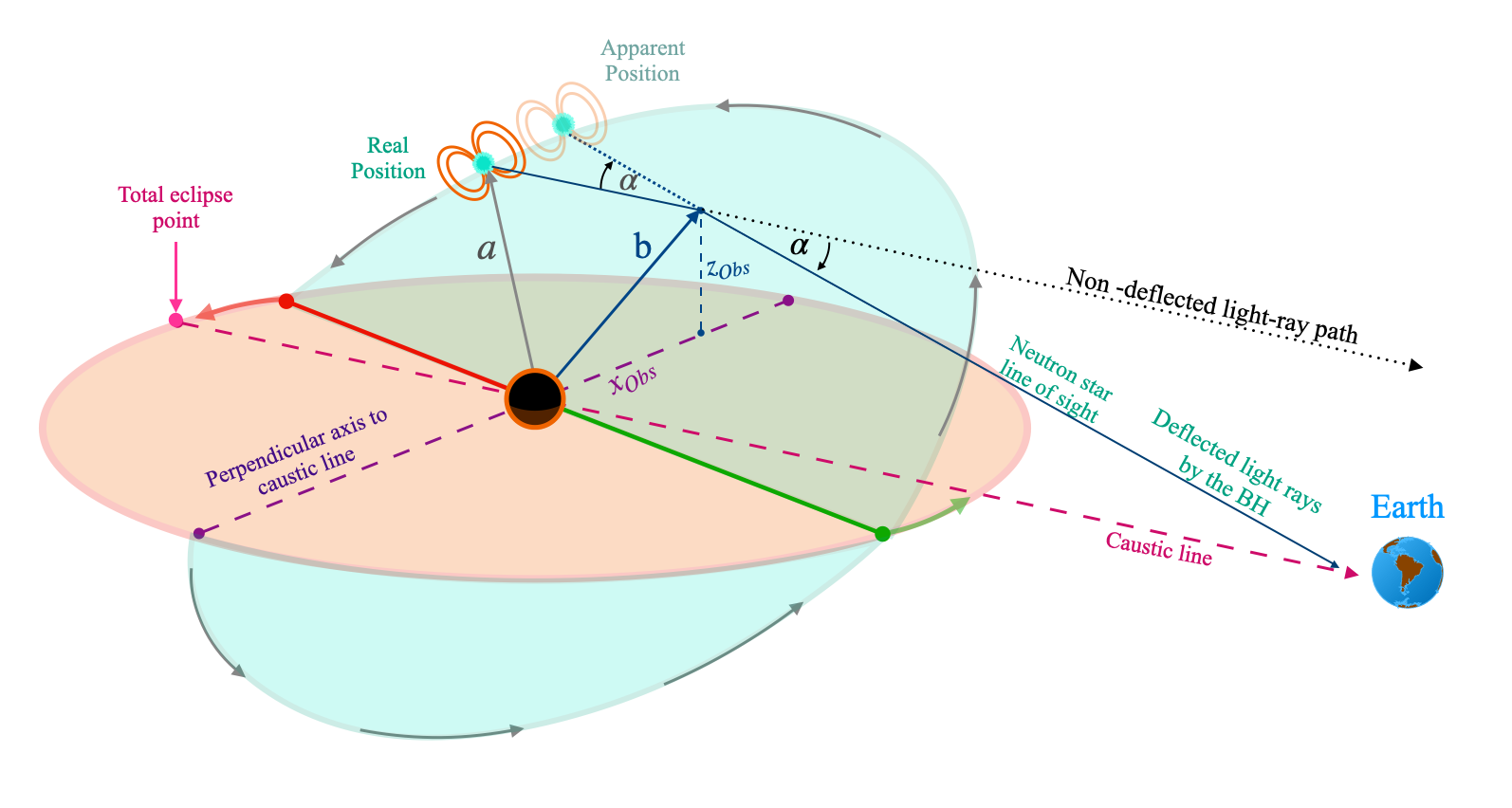}
        \caption{\small{ {Geometry used to compute the amplification by self-lensing. The caustic line is along the $y$-axis, and the projected source--lens separation is defined as $b=\sqrt{x_{\rm Obs}^2+z_{\rm Obs}^2}$.}}}
        \label{fig:AmplificationModel2}
    \end{subfigure}
    
 \caption{Schematic view of the NS–BH system and its self-lensing mechanism. 
Panel (a) shows the orbital configuration, (b) the gravitational lensing process producing a primary and a delayed secondary pulse, and (c) the geometry for burst amplification.}
    \label{fig:combined}
\end{figure}

\subsubsection{Activity Window}\label{subsec:aw}

The activity window is roughly generated in our model when the amplification $\mathcal{A}_T$ {reaches a given reference threshold where the pulses are significantly amplified}. This defines a threshold projected distance $b_T$ that denotes the transition from where the NS pulses can be detected. It can be inferred from Eq. (\ref{A_liebes}) by assuming that $8r_sa/b^2 \gg 1$, obtaining:
\begin{equation}\label{bT}
\frac{b_T}{r_s}\sim \frac{1}{2\mathcal{A}_T}\sqrt{\frac{8a}{r_s}},
\end{equation}
{The term $\mathcal{A}_T$ is a free parameter, but we set a reference threshold value of $\mathcal{A}_T\sim10^2$ motived by the energy gap between the bright FRB-like events detected from the Galactic magnetar \object{SGR~1935+2154} and the faint bursts of \object{FRB~20121102A}.}

{For keplerian orbits, we consider the most favourable configuration, with the orbit lying in the BH equatorial plane and this plane viewed edge-on. Here, $i$ denotes the inclination of the orbital plane relative to the BH equatorial plane, so this reference configuration corresponds to $i=0^\circ$.}
{In this case, the orbital plane remains fixed and the lensing configuration repeats once per orbit, so we identify $\tau_{\rm FRB}=\tau_{\rm Orb}$.}
The extension of the activity window $\tau_{\rm aw}^{\rm KP}$ can then be derived by considering that the NS crosses a projected distance $2b_T$ behind the BH with an approximately constant velocity $v_{\rm KP}=2\pi a/\tau_{\rm Orb}$. Then,
\begin{equation}\label{aw_KP}
\tau_{\mathrm{aw}}^{\mathrm{KP}} \sim \frac{1}{2\pi\mathcal{A}_T}\sqrt{\frac{r_s}{a}} \tau_{\mathrm{FRB}}.
\end{equation}
Here, the orbital period corresponds to $\tau_{\mathrm{FRB}}$. The amplification along the orbit $\mathcal{A}^{\mathrm{KP}}_{\mathrm{aw}}(t)$ can be estimated by considering that the projected distance $b(t)$ in Eq. (\ref{A_liebes}) is described as the projection of a uniform circular motion.  
\begin{equation}\label{Max_Amp_phase}
\mathcal{A}^{\mathrm{KP}}_{\mathrm{aw}}(t) \sim \frac{1}{2\pi}\sqrt{(2\pi)^2+ \frac{8r_s}{a (1-2\frac{t}{\tau_{\mathrm{FRB}}})^2}},
\end{equation}
where we must remark that this amplification is valid until $b\sim 3 R_{\mathrm{NS}}$, then the amplification is around $\mathcal{A}_{\max}$ given by Eq. (\ref{A_max}). We can estimate an upper limit for the rate of bursts with enough amplification to be detected. This approximate upper limit is computed by assuming that the neutron star can emit a number of pulses equal to the ratio between the time it takes to cross a distance $2b_T$ behind the BH and the {emission} period of the star $P_{\mathrm{NS}}$.
\begin{equation}\label{rates_KP}
 N^{\mathrm{KP}}(t)\sim  \frac{b_T}{2\pi aP_{\mathrm{NS}}}\tau_{\mathrm{FRB}}.
\end{equation}
 On the other hand, when considering {post-keplerian orbits} under LTp, the activity window is extended {when the orbit gets} an inclination $i$ {with respect to the fixed BH equatorial plane, which is viewed edge-on}. Then, in this scenario, the threshold distance $b_T$ is obtained by considering the projection of the inclined orbit on the plane where the nodes move, leading to 
\begin{equation}\label{bT_LT}
\frac{b_T}{r_s}\sim \frac{1}{2\mathcal{A}_T\sin(i)}\sqrt{\frac{8a}{r_s}},
\end{equation}
Following the idea applied to the Keplerian orbit in Eq. (\ref{aw_KP}), it is straightforward to show that the time span of the activity window $\tau_{\mathrm{aw}}^{\mathrm{LT}}$ can be computed as{
\begin{equation}\label{aw_LT}
\tau_{aw}^{LT} \sim  \frac{1}{\mathcal{A}_T\pi \sin(i)}\sqrt{\frac{8r_s}{a}} \tau_{\mathrm{FRB}},
\end{equation}}
where we {recall} that in this case the FRB cycle $\tau_{\mathrm{FRB}}$  is half of  the Lense-Thirring time-scale $\tau_{\mathrm{LT}}$, see {Eq. (\ref{T_FRB_LT})}. For the case of \object{FRB 20121102A} and considering $\mathcal{A}_T=10^2$ and a distance $r_s/a\sim 1/200$, we need an inclination $i\sim (5-8)^{\circ}\times 10^{-2}$ to obtain $ \tau_{\mathrm{aw}}^{\mathrm{LT}}\sim 0.25 \tau_{\mathrm{LT}} = 0.5 \tau_{\mathrm{FRB}}$, which corresponds roughly to the activity window of \object{FRB 20121102A}.
The maximum amplification that can be obtained along the orbital phase can be computed by considering that the NS must cross a distance $2b_T$, but $b_T$ changes on each orbit. At the beginning of the activity window $b_T$ is a short distance near the border of the BH, while at the middle of the activity window it diametrically crosses the BH. By setting this geometry we obtain
\begin{equation}\label{Max_Amp_phase_LT}
\mathcal{A}^{\mathrm{LT}}_{\mathrm{aw}}(t) \sim \frac{1}{2}\sqrt{\frac{2\pi}{\sin(i)}+ \frac{8r_s}{a (1-2\frac{t}{\tau_{\mathrm{FRB}}})^2}},
\end{equation}
Finally, the rates per day of the FRB follow the same idea as Eq. (\ref{rates_KP}) but for a $b_T$ that changes on each orbit
\begin{equation}\label{LT_rates}
 N(t)^{LT}\sim  \frac{2b_T}{\pi aP_{\mathrm{NS}}}\sqrt{1-\Big(1-\frac{2t}{\tau_{\mathrm{FRB}}}\Big)^{2}}\tau_{\mathrm{FRB}}.
\end{equation}
We remark that all this rate inference was made under the assumption that all pulses follow a strict periodicity without any contributions from spin precession or slow-down. 

\subsubsection{Pulses after amplification}\label{subsec:PaA}

The radio pulses that are exposed to the  physics in the vicinity of a BH experience secondary effects of lensing.
The most significant effect is the splitting of a burst into two images: a primary image and a secondary one. Throughout this work, we refer to the secondary component as the echo (see Fig. \ref{fig:magni}).
{This phenomenon can be identified observationally through the arrival-time offset between the primary pulse and its lensed echo. 
Following \citet{schneider}, it can be written as}{
\begin{equation}\label{Eq:image_delay}
\tau_{\mathrm{lens}}^{\mathrm{Spl}} =
\frac{4GM_{\mathrm{BH}}}{c^3}(1+z)\tau(\bar{b}),
\end{equation}}
where $4GM_{\mathrm{BH}}/c^3$ sets the characteristic relativistic time-delay scale {of the lens: for} BH masses of $10^{3{-}4}~M_{\odot}$, this scale is $\sim \mathcal{O}(10^{-2}{-}10^{-1}~\mathrm{s})$. The term $z$ denotes the redshift of the host galaxy. The normalized projected distance is defined as $\bar{b}=b/R_E$, with $R_E=\sqrt{2r_s a}$ the Einstein-ring radius. This expression assumes that the observer--BH distance is approximately equal to the observer--NS distance, and {uses the fact that the BH--NS separation equals the semi-major axis $a$ for the circular orbits considered here}.

The dimensionless function \( \tau(\bar{b}) \) in Eq. (\ref{Eq:image_delay}) depends on the proximity of the NS to the caustic line in its orbit and is given by
\begin{equation}\label{tau}
\tau(\bar{b}) = \frac{1}{2}\bar{b}\sqrt{\bar{b}^2 + 4} + \ln\left( \frac{\sqrt{\bar{b}^2 + 4} + \bar{b}}{\sqrt{\bar{b}^2 + 4} - \bar{b}} \right).
\end{equation}
Assuming that most of the amplification occurs when the NS is directly behind the BH, we can assume that  $\bar{b}\lesssim 1$. We estimate an order of magnitude in Appendix \ref{app:lambda}, leading to  {($\tau_{\mathrm{lens}}^{\mathrm{Spl}} \sim \mathcal{O}(10^{-2}-10^{-3} \mathrm{s}\,)$}.

Another key feature of the split pulse is the intensity ratio between the two components. Following \citet{schneider}, the luminosity ratio $R_L$ between the brighter main pulse ($L_{\mathrm{main}}$) and the fainter echo ($L_{\mathrm{echo}}$) can be computed as:
\begin{equation}
R_L\equiv \frac{L_{\mathrm{main}}}{L_{\mathrm{echo}}} = \left[\frac{\sqrt{\bar{b}^2 + 4} + \bar{b}}{\sqrt{\bar{b}^2 + 4} - \bar{b}}\right]^2.
\label{eq:luminosity_ratio}
\end{equation}
{By taking the same assumptions made for the delay between pulses, the ratio can be rewritten in an approximate form as $R_L\approx1+2\bar b$, where they are equal in intensity when the NS and the BH are fully aligned ($\bar b=0$), as the pulses merge into the Einstein ring.}

These effects can be incorporated into the burst simulation by including the echo at a time delay $\Delta t$, as given by Eq. (\ref{Eq:image_delay}), following the main burst. The intensities of the two components are adjusted using Eq. (\ref{eq:luminosity_ratio}), under the assumption that their sum satisfies energy conservation: $L_{\mathrm{main}} + L_{\mathrm{echo}} = L_{\mathrm{original}}$, where $L_{\mathrm{original}}$ is the luminosity of the burst after amplification that does not consider burst splitting. 

{In addition to the lensing-induced image delay discussed above, the arrival times of the pulses can be affected by the standard timing corrections used in compact binary pulsar systems, namely the Rømer, Einstein, and orbital Shapiro delays. These effects introduce phase-dependent corrections to the observed pulse arrival times as the NS moves along its orbit. In the simulations, we include these timing corrections when constructing the waiting-time distribution, following the expressions summarized in Appendix~\ref{app:timing_corrections}.}

\subsection{Code structure}\label{subsec:code}

The code takes as input the FRB cycle period $\tau_{\mathrm{FRB}}$, the BH spin $\chi$ and mass $M_{\mathrm{BH}}$, the orbital inclination $i$ and eccentricity $e$, and the NS {cadence} period. In our simulations, we set $\chi = 0.9$ {as a high-spin reference case} and adopt $e = 0$, as the orbital eccentricity is expected to decay to zero due to gravitational-wave emission.

The orbital motion is generated as a sequence of discrete points computed from the equations described in Sec.~\ref{subsec:equations}. {These points are equally spaced by an assumed emission period $P_{\rm NS}$, representing potential radio emission events. Here, $P_{\rm NS}$ specifies the pulse cadence rather than necessarily the NS rotation period; the directional selection associated with the emission beam is not explicitly modelled.}

For each orbital position lying behind the BH, the code evaluates the amplification factor using the lensing equations from Sec.~\ref{subsec:amplification}. The amplification depends solely on the relative position of the NS with respect to the BH. Each lensed event is then assigned an intrinsic pulse energy drawn from a log-normal distribution \citep{VeraCasanova2025}, representing the emission energy prior to lensing. This energy is subsequently multiplied by the computed amplification factor.

Corrections due to burst splitting are then applied to the amplified bursts, following the prescriptions in Sec.~\ref{subsec:PaA}. These corrections depend exclusively on the orbital position of the NS at the time of emission and on the redshift of the host galaxy. Focusing on \object{FRB~20121102A}, we adopt $z = 0.19$ \citep{Tendulkar2017}.

{To analyse the resulting energy distribution and its modulation along the orbital phase, the code compares the energies of all lensed bursts with the original log-normal input distribution. Finally, it computes the waiting times by taking the differences between the arrival times of consecutive observable bursts, incorporating the time-aberration corrections described in Appendix~\ref{app:timing_corrections} and accounting for the adopted redshift of the host galaxy.}

\section{Results and discussion}\label{sec:Results}
\subsection{Keplerian orbits}\label{subsec:Kepler}
{We begin by analysing the most favourable aligned Keplerian orbit with zero inclination ($i = 0$), assuming that the orbital period matches the full cycle of the FRB ($\tau_{\mathrm{Orb}} = \tau_{\mathrm{FRB}}$). Other inclinations are also valid Keplerian orbits, but they lead to poorer projected alignments and therefore to lower lensing probabilities.}

The corresponding activity window for this orbital configuration is shown in Fig.~\ref{fig:Kp_ampli}, and its duration can be estimated using Eq.~(\ref{aw_KP}) by defining it as the orbital range where the pulses exhibit an amplification factor $\mathcal{A}_T \geq 100$. 

From Eq.~(\ref{aw_KP}) it is possible to observe that this  activity window depends on the inverse of the system’s scale, $r_s/a \sim \mathcal{O}(10^{-4}$–$10^{-5})$, and on the inverse of the amplification threshold, $\mathcal{A}_T^{-1} \sim 10^{-2}$. Thus, the predicted activity window is $\tau_{\mathrm{aw}}^{\mathrm{KP}} \sim \mathcal{O}(0.5\,\mathrm{hours})$. This timescale is several orders of magnitude shorter than the observed activity windows of prFRBs, which span $\sim 30$–$50\%$ of their periods (e.g., $\sim 80$ days in the case of \object{FRB 20121102A}), indicating that this scenario alone cannot account for the observations.

Additionally, the number of detectable bursts depends on the time the NS takes to cross a distance of $2b_T$, relative to its rotation period $\tau_{\mathrm{NS}}$. For a young magnetar with periods ranging from a few seconds to sub-second scales, this results in only a sparse chain of detectable bursts—ranging from hundreds to thousands—distributed throughout the orbital cycle, as predicted by Eq. (\ref{rates_KP}).

In exploring the possibility of extending the activity window in purely Keplerian orbits, we examine compact system configurations. Varying the system scale ($a/r_s$) reveals an additional limitation. As the NS orbits closer to the BH, the effective path length over which lensing can occur increases, potentially broadening the duration of the eclipsing region. However, the orbital velocity grows even more rapidly, leading to a net decrease in the total lensing time. Simultaneously, the lifetime of such tight systems becomes significantly shorter because of strong gravitational wave emission, rendering these configurations unlikely to account for the observed behavior of prFRBs.

This constraint becomes particularly severe for NSs in tight orbits around the BH. In such cases, the eclipse duration $\tau_{\mathrm{eclipse}}$ may become shorter than the NS's {cadence} period $P_{\mathrm{NS}}$. When $\tau_{\mathrm{eclipse}} < P_{\mathrm{NS}}$, the probability of a burst being emitted precisely during the alignment needed for strong lensing drops significantly. As a result, the specific lensing geometry required for detectable FRB production becomes increasingly unlikely in compact Keplerian orbital configurations. Therefore, we conclude that purely Keplerian orbits are unlikely to account for the observed activity windows of known prFRBs.

\begin{figure}[htbp]
    \centering
    \includegraphics[width=1\linewidth]{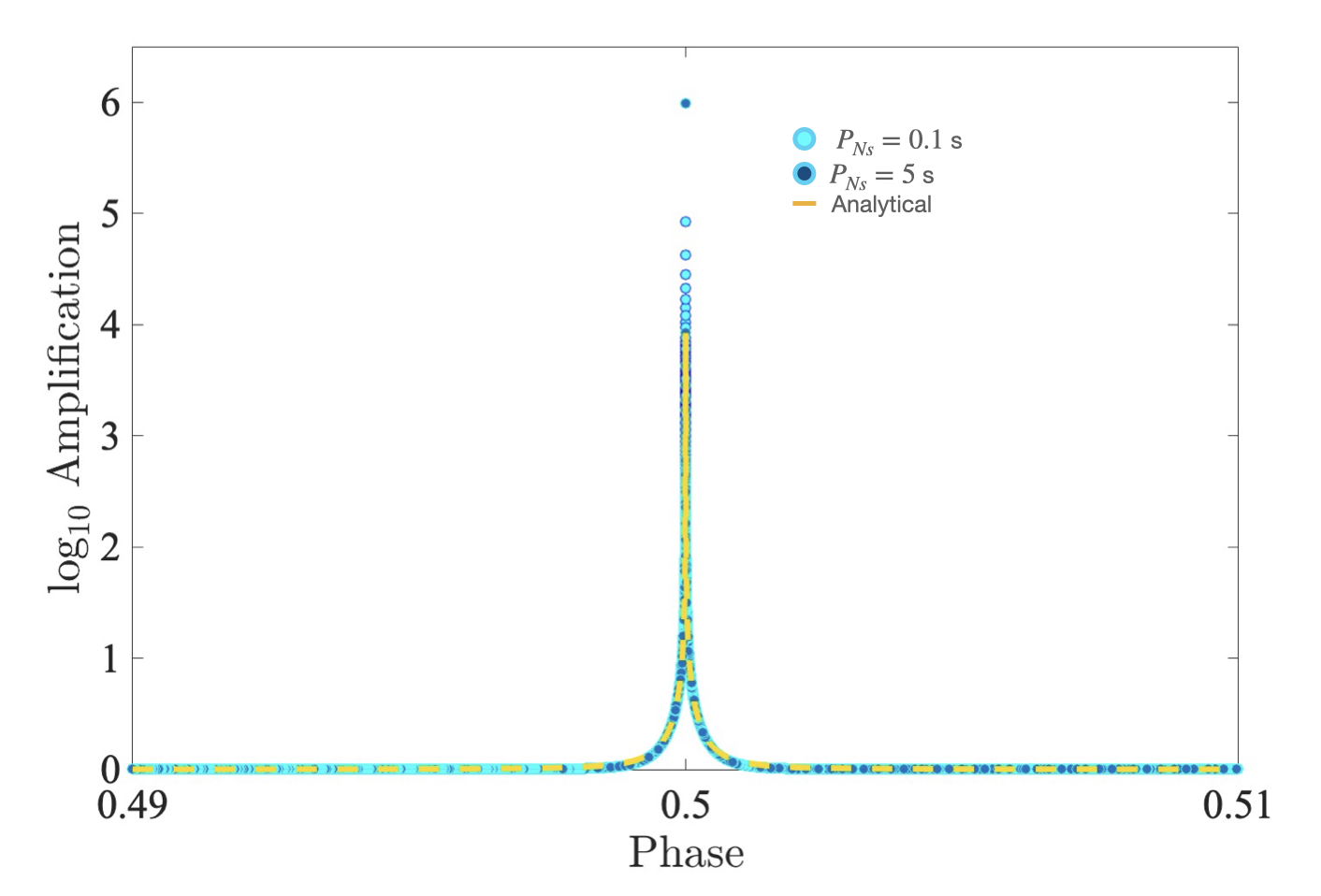}
    \caption{
        Mapping of the amplification of the NS signal as a function of orbital phase along a Keplerian orbit. This simulation assumes a BH mass of \( M_{\mathrm{BH}} = 10^{3.5}\, M_\odot \) and an orbital period of \( \tau_{\mathrm{Orb}} = 160 \) days. Blue and cyan dots represent the simulated amplification for NSs with {cadence} periods \( P_{\mathrm{NS}} = 5 \)~s and \( P_{\mathrm{NS}} = 0.1 \)~s, respectively. The yellow dashed line shows the analytical prediction from Eq.~(\ref{Max_Amp_phase}). {Beyond the plotted x-axis range, the amplification approaches unity}.
    }\label{fig:Kp_ampli} 
\end{figure}

\subsection{Precessing orbits}\label{subsec:LTOrbits}

When including Lense–Thirring precession (LTp), the orbital inclination and frame dragging are the key to extending the activity window, as shown in Eq.~(\ref{aw_LT}). In these configurations, the full FRB cycle ($\tau_{\mathrm{FRB}}$) corresponds to half the Lense–Thirring period, i.e., $\tau_{\mathrm{LT}} = 2\tau_{\mathrm{FRB}}$. One cycle occurs when the ascending node passes behind the BH, and the other when the descending node does so. In this scenario, the observed FRB cycle sets the LTp timescale, which in turn constrains the semi-major axis $a$ through Eq.~(\ref{LTpres}), once the BH mass and spin are specified.

The observable features of these systems—such as the duration of the activity window, the amplification, and the burst rate—depend sensitively on the system's scale, characterized by the ratio $a/r_s$. To investigate this dependence, we explore how the system geometry varies with different BH masses while keeping $\tau_{\mathrm{FRB}}$ fixed. This allows us to assess the impact on additional key parameters such as the system's lifespan, orbital period, and the maximum achievable amplification.

Our results show that, {for a fixed FRB cycle, BH spin, and orbital eccentricity}, the ratio $a/r_s$ decreases as the BH mass increases (see Fig.~\ref{fig:system}). Specifically, {the configurations reproducing the same FRB cycle satisfy $a/r_s \propto M_{\rm BH}^{-1/3}$. Thus, the required orbital separation decreases in units of the Schwarzschild radius, although the physical separation increases as $a \propto M_{\rm BH}^{2/3}$}. {Along this family of configurations}, the orbital period ranges from several minutes in the case of stellar-mass BHs to several days for supermassive ones.

\begin{figure*}[t]
    \centering
    \begin{subfigure}[t]{0.48\textwidth}
        \centering
        \includegraphics[width=\linewidth]{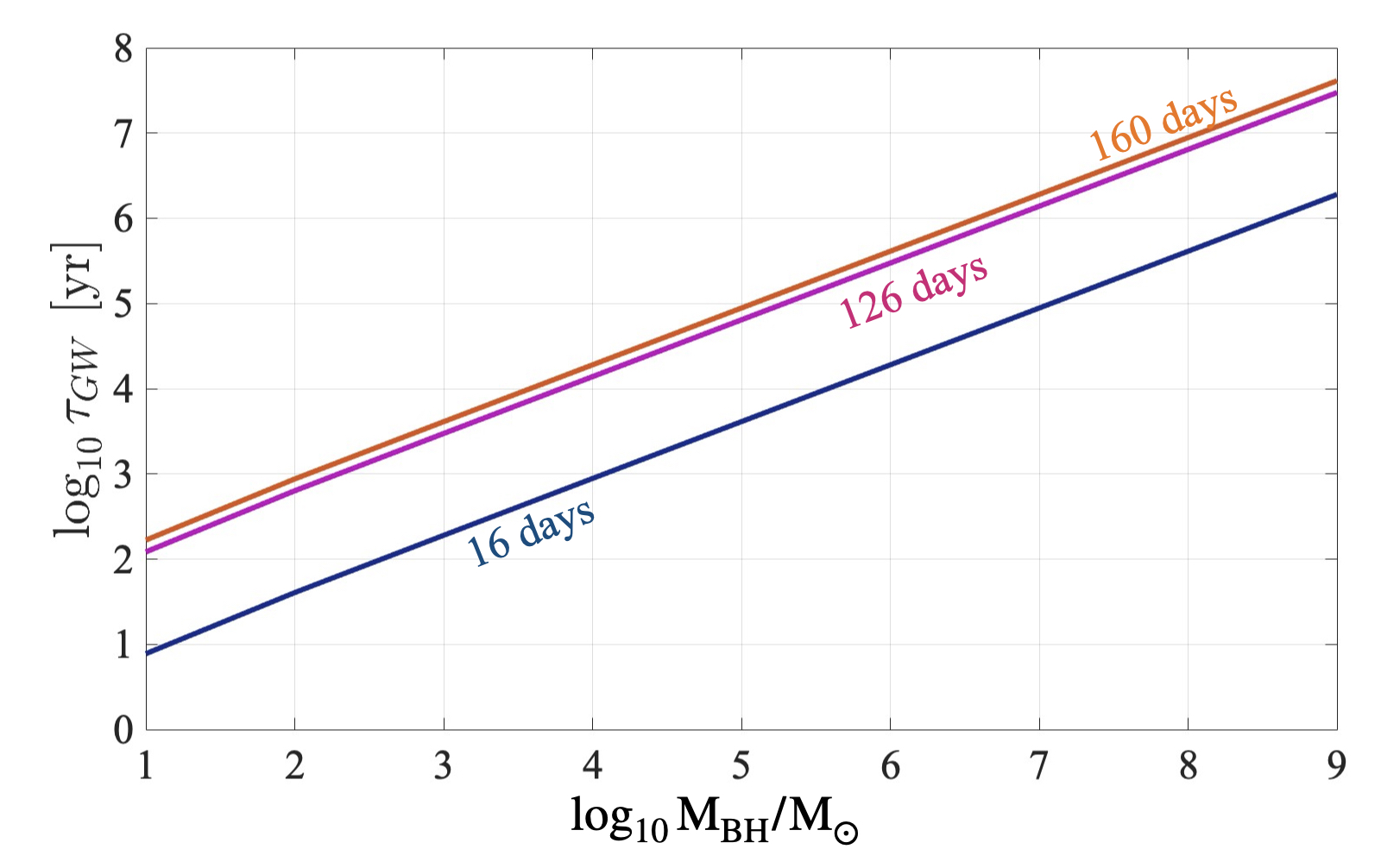}
    \end{subfigure}
    \hfill
    \begin{subfigure}[t]{0.48\textwidth}
        \centering
        \includegraphics[width=\linewidth]{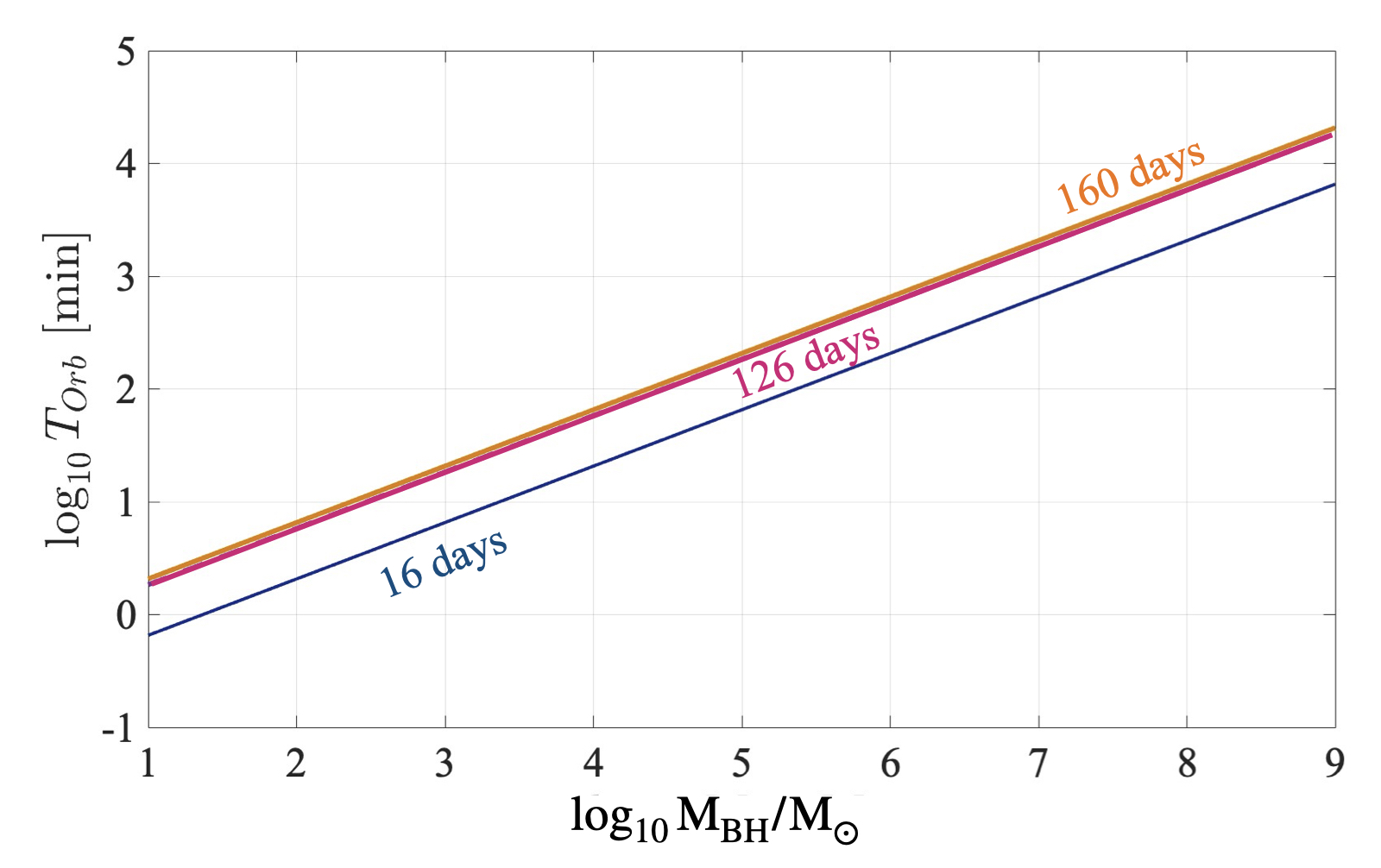}
    \end{subfigure}

    \vspace{0.5em}

    \begin{subfigure}[t]{0.48\textwidth}
        \centering
        \includegraphics[width=\linewidth]{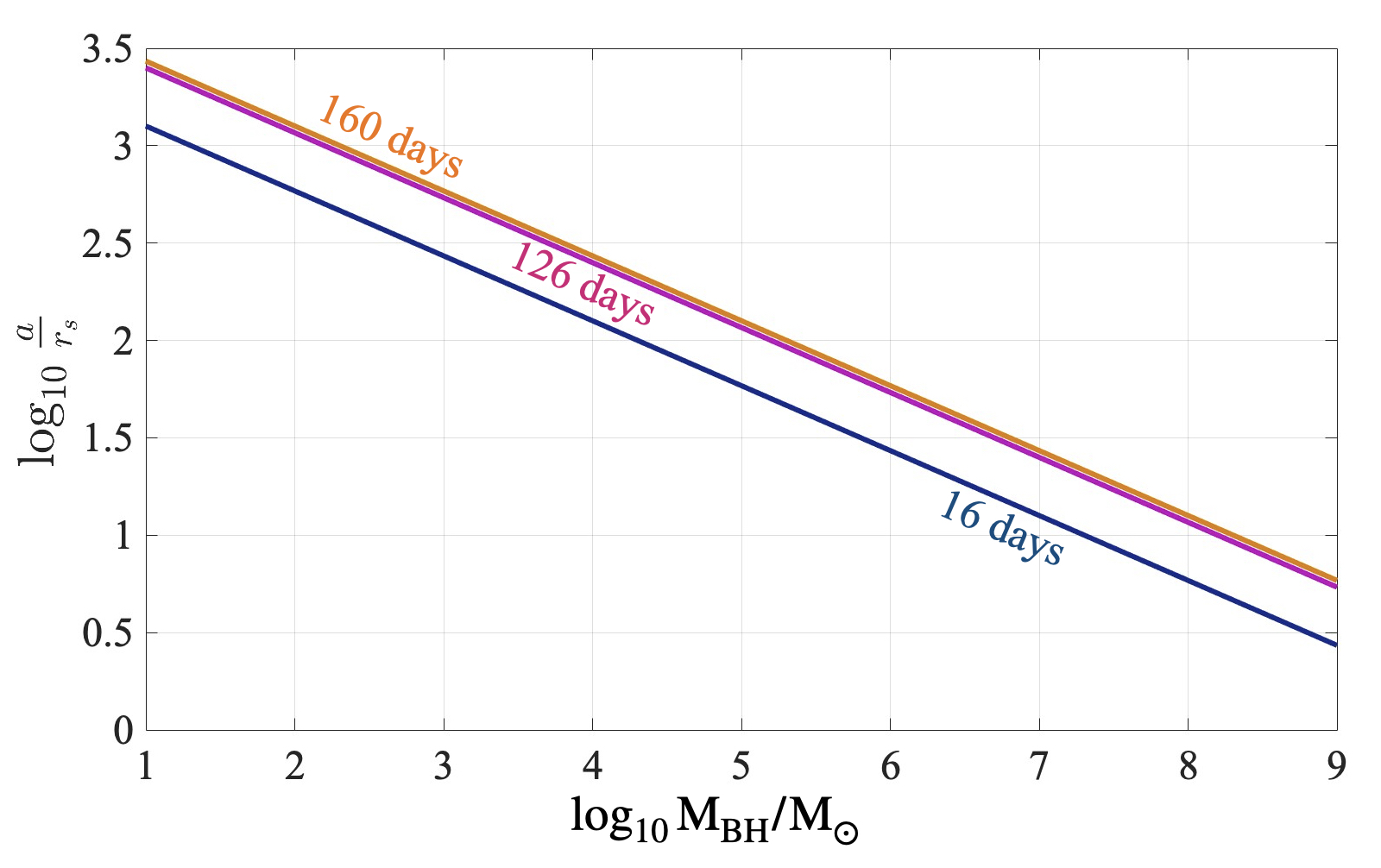}
    \end{subfigure}
    \hfill
    \begin{subfigure}[t]{0.48\textwidth}
        \centering
        \includegraphics[width=\linewidth]{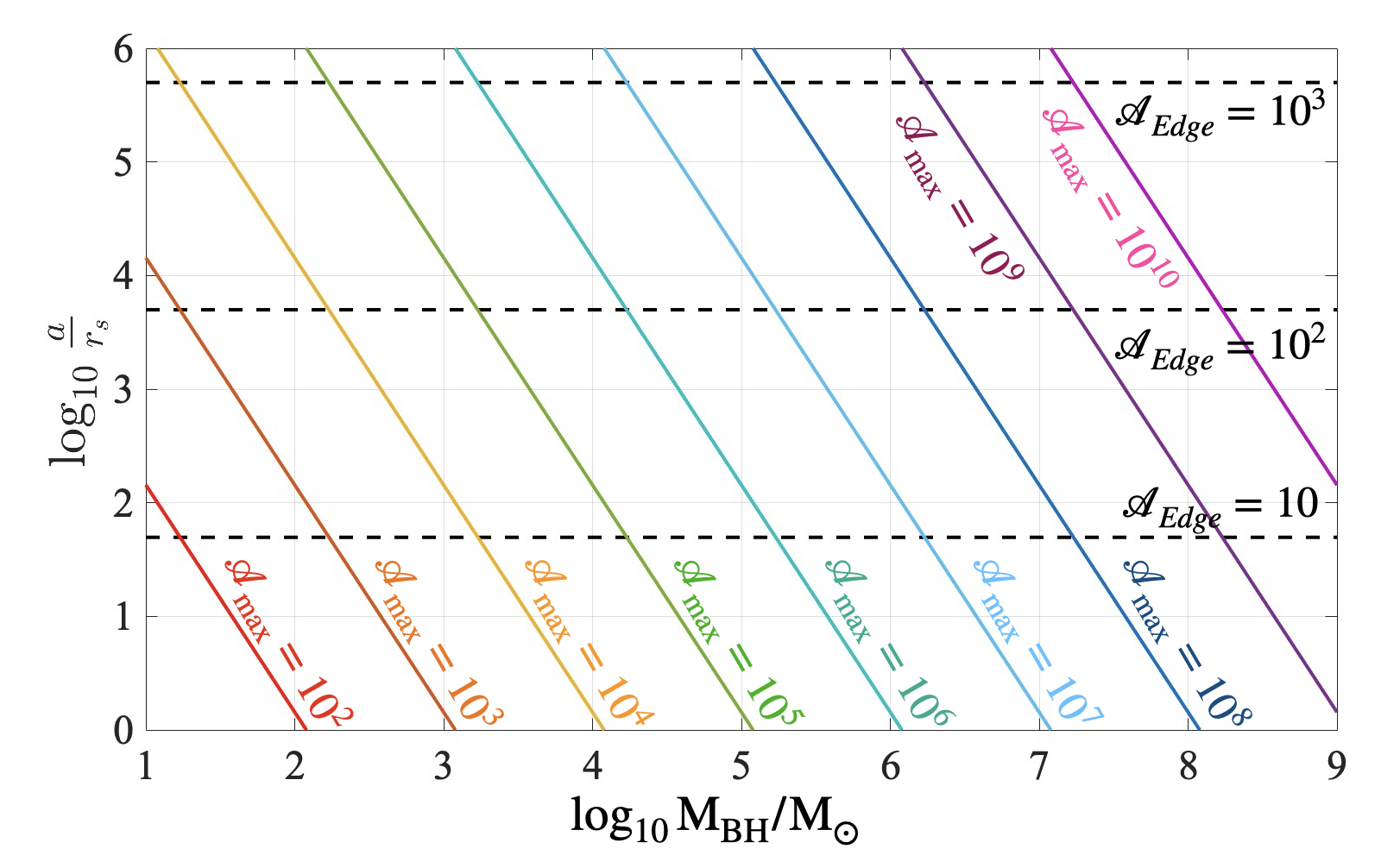}
    \end{subfigure}

    \caption{\textbf{Top left.} Inspiral time of the system due to gravitational wave emission as a function of the BH mass. Different curves correspond to varying FRB periods: orange for 160 days, purple for 126 days, and blue for 16 days. \textbf{Top right.} Orbital period for each BH mass, considering the three aforementioned FRB cycles. \textbf{Bottom left.} Relationship between the BH mass and the orbital radius (system's scale), indicating the necessary configuration to reproduce FRB periods. \textbf{Bottom right.} Maximum amplification of the FRB signal due to lensing, depicted by diagonal color lines, as a function of the BH mass and the system's scale. Horizontal dashed lines show the amplification at the edge of the BH.}
    \label{fig:system}
\end{figure*}

The inspiral timescale for such systems is shown in Fig.~\ref{fig:system}. According to Eq. (\ref{lifespan}), the gravitational-wave timescale scales as $\tau_{\mathrm{GW}} \propto a^4/M_{\mathrm{BH}}^2$, while the constraint $a \propto M_{\mathrm{BH}}^{2/3}$ (required to maintain $\tau_{\mathrm{LT}} = 2\tau_{\mathrm{FRB}}$) implies that $\tau_{\mathrm{GW}} \propto M_{\mathrm{BH}}^{2/3}$. Counterintuitively, this means that systems involving more massive BHs are longer-lived. As a consequence, under this model, stellar-mass BHs become unfavorable candidates for hosting long-lived prFRB systems.

As it was mentioned before, the maximum amplification due to self-lensing occurs when the NS crosses the caustic line. This value, given by Eq.~(\ref{A_max}), depends on the system scale. As shown in Fig.~\ref{fig:system} {(bottom right panel)}, the maximum amplification is represented by the diagonal lines: it increases with both the BH mass and the orbital separation. The horizontal dashed lines represent the amplification evaluated at the BH’s edge, obtained by setting $b = r_s$ in Eq.~(\ref{A_liebes}). Expressing this result in units of $r_s$ removes the explicit dependence on the BH mass, making it a function solely of the system scale $a/r_s$. Therefore, the figure provides an estimate of the expected amplification range—from a relative minimum at the BH’s edge to the maximum at the caustic—for various masses and orbital distances.

In Fig.~\ref{fig:AW_Amplification}, we present a representative example of the amplification window for FRB~20121102A, considering orbits with LTp. As in the previous section, we define the activity window as the phase interval during which the pulses—assumed to follow a log-normal energy distribution—are sufficiently amplified such that their average energy exceeds the detection threshold ($\mathcal{A}_T = 100$). The overall extent of the activity window is well described by the envelope curve shown in Fig.~\ref{fig:AW_Amplification}, given analytically by Eq.~(\ref{Max_Amp_phase_LT}), which depends only on the orbital inclination and the system's scale, and is therefore relatively weakly sensitive to the intrinsic pulse energy or the NS's {cadence} period. 

The characteristic shape of the amplification curve, both in Eq. (\ref{Max_Amp_phase_LT}) and in the simulations shown in Fig. \ref{fig:AW_Amplification}, arises from the diagonal sweeps of the NS behind the BH. At the beginning of the activity window, the NS passes near the BH’s limb and spends only a short time behind it, resulting in weakly amplified or marginally detectable bursts. As the ascending node (or descending node) approaches the caustic line, the NS traverses a longer projected path behind the BH, maximizing the amplification near the center of the activity window, where the NS crosses the BH nearly diametrically. After this point, the crossing distance and amplification decrease as the node moves away from the caustic, ending the activity window when the NS again approaches the outer BH edge.

Another direct consequence from gradually sweeping of the NS across the BH is the extension of the activity window. Unlike in a purely Keplerian orbit, where the activity window opens only when the projected distance from the NS to the BH approaches the Schwarzschild radius, here the activity begins when the ascending node (or descending node) approaches a projected distance of $b\sim r_s/\sin i$. For small inclination angles, $i \sim 10^{-2}-10^{-1}$ degrees, the activity window can span a significant fraction of the Lense-Thirring precession period.

Furthermore, this analysis reveals that for a given BH mass, the observed activity fraction of any prFRB can be reproduced by appropriately adjusting the inclination angle $i$. This is demonstrated in Fig.~\ref{fig:AWps}, where values around $i \sim \mathcal{O}(10^{-2} -10^{-1}~^{\circ})$ {yield the observed activity fraction.} The yellow banana-shaped region in the figure shows the combinations of $i$ and $M_{\mathrm{BH}}$ that yield a ratio between the activity window and the full cycle of approximately $50\%$, consistent with observations of FRB~20121102A. The dashed red line corresponds to the theoretical prediction from Eq.~(\ref{aw_LT}). Parameter combinations above this region (in blue) underpredict  the activity window duration, while those below it overpredict it.

As shown in Fig.~\ref{fig:counts}, the number of detectable events per day in LTp orbits within the activity window increases significantly compared to the purely Keplerian case. This enhancement occurs because the NS remains much closer to the BH, and its total motion results from the combination of a faster diagonal Keplerian crossing and the additional frame dragging induced by the BH’s rotation. As a consequence, the number of eclipses $N_E$ increases from one per orbit (in the Keplerian case) to approximately $N_E \sim \frac{\tau_{\mathrm{LT}}}{2\tau_{\mathrm{Orb}}} \sim \mathcal{O}(10^4)$, assuming $\tau_{\mathrm{LT}} = 2\tau_{\mathrm{FRB}}$.


The assumed BH spin affects the orbital scale required to reproduce a fixed FRB cycle. {In our fiducial simulations, we adopt $\chi=0.9$ as a high-spin reference case, rather than as a representative value for the IMBH population. Successive mergers with lower-mass companions can lead to substantially lower spins \citep{Kritos2024}, while sustained coherent accretion can spin up the BH \citep{Volonteri2005,Berti2008}. However, a high spin is not required to reproduce the adopted activity cycle and activity fraction: Table~\ref{tab:spin_dependence} illustrates suitable parameter combinations for $0.1\leq\chi\leq0.9$ within our analytical approximation.}

For fixed $M_{\rm BH}$ and $\tau_{\rm FRB}$, imposing $\tau_{\rm LT}=2\tau_{\rm FRB}$ gives $a\propto\chi^{1/3}$. The orbital period scales as $\tau_{\rm orb}\propto\chi^{1/2}$, while the maximum amplification depends only weakly on spin, $A_{\rm max}\propto\chi^{1/6}$. For example, reducing the spin from $\chi=0.9$ to $\chi=0.3$ decreases the orbital period by only a factor $\simeq0.58$, and the maximum amplification by $\simeq0.83$. Therefore, moderate changes in the assumed spin do not qualitatively modify the main results of the model.


The predicted burst rate is highly sensitive to the NS {cadence} period, as expressed in Eq.~(\ref{LT_rates}). This dependence is clearly illustrated in Fig.~\ref{fig:counts}: the shorter the NS period, the higher the number of potentially detectable events. Consequently, observational constraints on the burst rate can be directly translated into upper limits on the spin period, yielding values ranging from a few seconds down to sub-second scales.

The potential detection of hundreds of bursts per day is roughly consistent with the rates reported by \citet{EspinozaDupouy2025}. However, it is important to note that our model simplifies the emission pattern to a strictly periodic one, while in reality the emission of strong bursts may follow a more Poissonian or stochastic behavior, modifying the waiting-time distribution and potentially explaining both the observed clustering of bursts within the activity window and the intervals with no detections.

Additionally, magnetars may undergo phases of enhanced activity, producing more powerful and recurrent bursts than average. Propagation effects in the intervening medium—such as dispersion, scattering, absorption, and interstellar scintillation—could also modulate or even obscure the signal, reducing the likelihood of detection at certain epochs.



{So far, we have defined a reference activity window using the nominal amplification threshold $\mathcal{A}_T=100$, which determines the projected distance threshold $b_T$ in Eq.~(\ref{bT}). This definition isolates the geometrical dependence of the amplification window without specifying the width of the intrinsic pulse-energy distribution. For the energy distribution, however, we select individual pulses whose lensed energies satisfy $E_{\rm lensed}=\mathcal{A}E_{\rm int}\geq10^2\langle E_{\rm int}\rangle$, where $\langle E_{\rm int}\rangle$ is the mean intrinsic pulse energy. Thus, the amplification threshold and the energy-selection criterion coincide only for pulses with the mean intrinsic energy.}

{Intrinsically brighter pulses can therefore remain detectable outside the reference amplification window, whereas weaker pulses may fall below the energy threshold even within it. For example, a pulse ten times more energetic than the intrinsic mean requires an amplification of only $\mathcal{A}\geq10$, corresponding approximately to a projected distance threshold $b_T^{\rm eff}\sim10b_T$ in the regime where $\mathcal{A}\propto b^{-1}$. The resulting detectable activity window therefore depends on the intrinsic energy distribution as well as the orbital geometry. A broader energy distribution can extend the detectable activity beyond the reference window, potentially requiring a larger orbital inclination to reproduce the same observed activity fraction. The magnitude of this adjustment, and its effect on the detected burst rate, depend on the fraction of pulses exceeding the energy threshold.}

\begin{figure}
    \centering
    \includegraphics[width=1\linewidth]{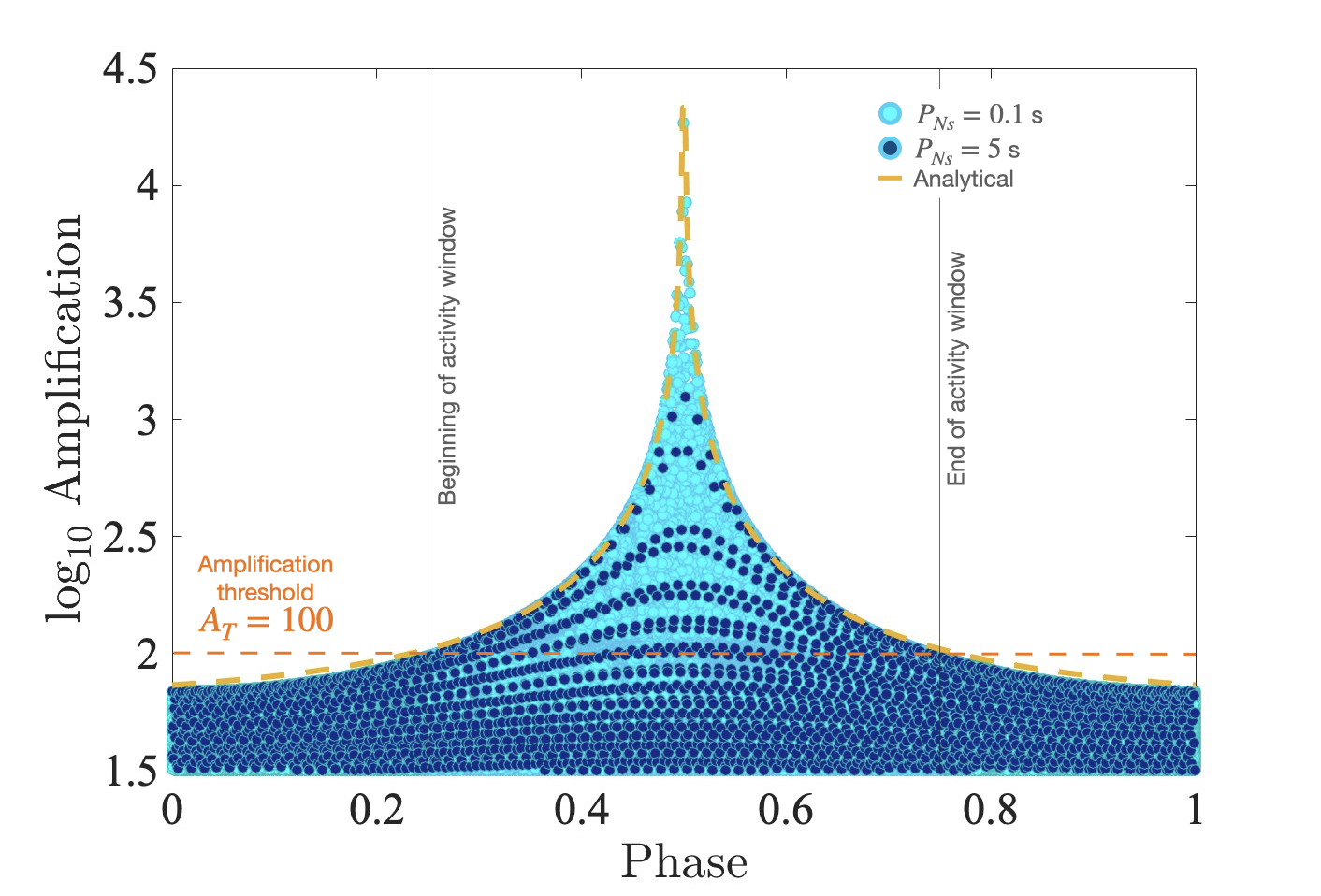}
    \caption{ Mapping of the gravitational lensing amplification of individual bursts across an entire FRB cycle.
        Each dot represents the amplification of a single pulse from the NS.
        Cyan dots correspond to simulations for an NS with a {emission cadence} period of $0.1$ s, while blue dots represent an NS with a {cadence} period of $5$ s. Parameters for these simulations are motivated by FRB~20121102A, considering an orbital cycle of $160$ days and an $80$-day activity window.
        Lensing effects were computed for a BH mass $M_{\mathrm{BH}}=10^{3.5}\,M_{\odot}$ and an orbital inclination of $i=0.05^\circ$. The activity window specifically indicates orbital phases where bursts experience amplification factors {$\mathcal{A}\gtrsim 100$}.}
    \label{fig:AW_Amplification}
\end{figure}

\begin{figure}
    \centering
    \includegraphics[width=1\linewidth]{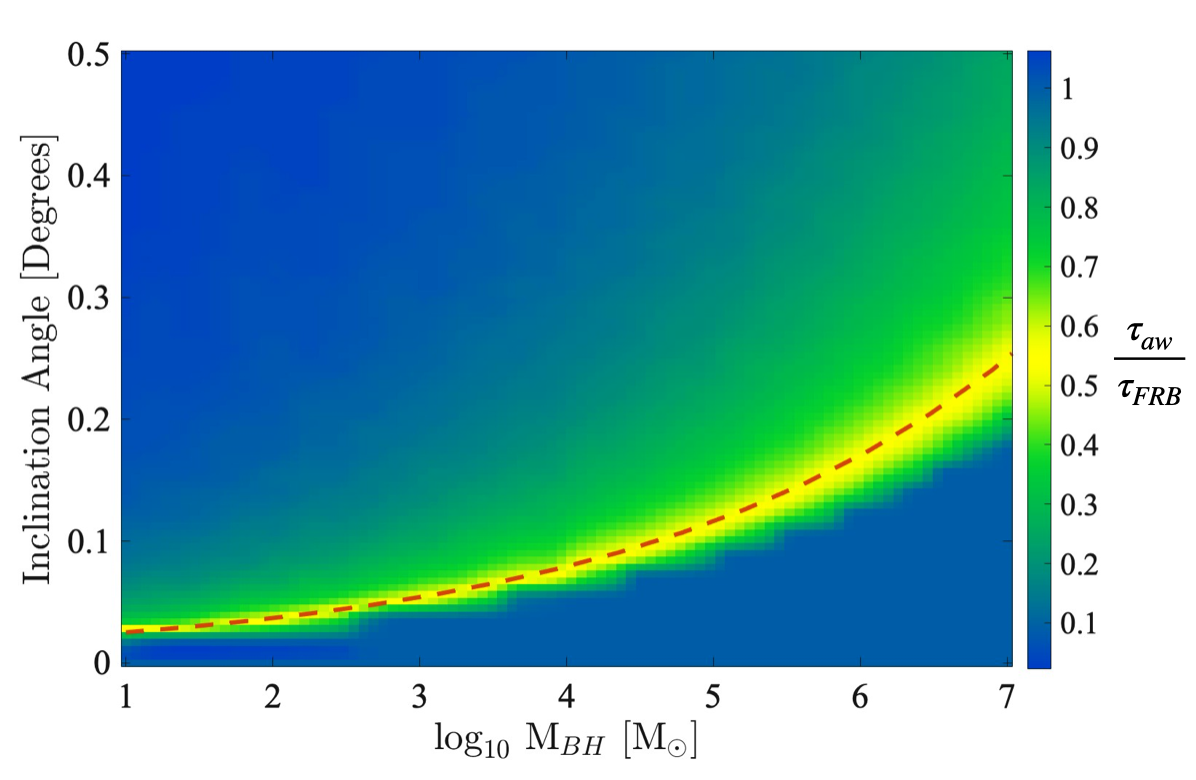}
    \caption{ Color map exploring the orbital inclination angle $i$ vs BH mass $M_{\mathrm{BH}}$ to reproduce the activity window duration of FRB~20121102A. The banana-shaped yellow region indicates that the system would yield an activity window $\tau_{\mathrm{aw}}$ of $\frac{\tau_{\mathrm{FRB}}}{\tau_{\mathrm{aw}}}\sim 50\%$, matching the observed value for FRB~20121102A. The blue region above the banana-shaped zone represents parameter combinations that underestimate the observed activity-window duration, while the blue region below overestimates it. The orange dashed line represents the theoretical prediction for the relation between BH mass and inclination angle.}
    \label{fig:AWps}
\end{figure}

\begin{figure}
    \centering
    \includegraphics[width=1\linewidth]{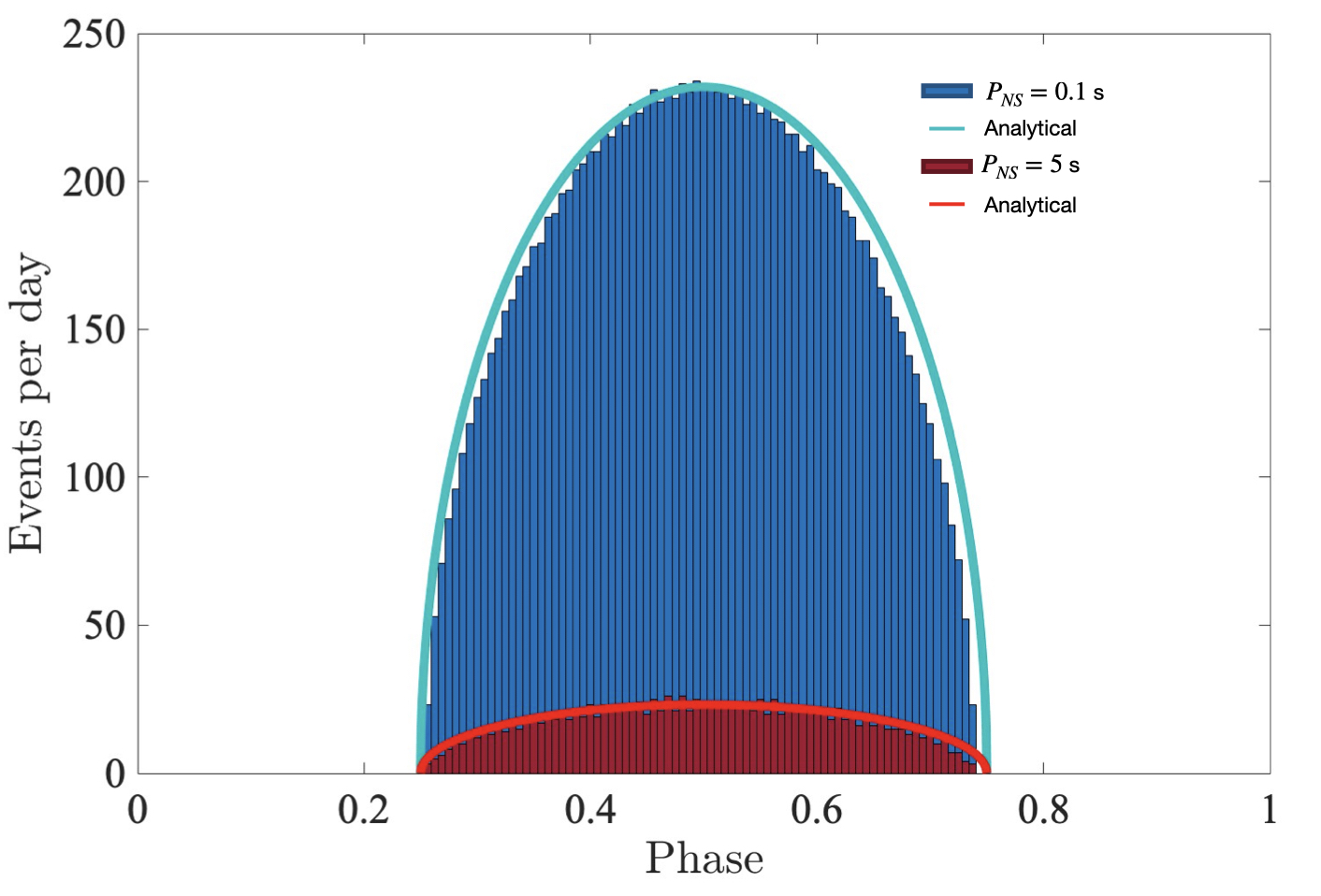}
    \caption{{Number of bursts expected per day within the activity window. The estimate is computed for \object{FRB~20121102A} with $M_{\rm BH} = 10^{3.5} M_{\odot}$ and orbital inclination $i = 0.05^\circ$. Red and blue bars show the predicted detection rates for a NS with a {cadence} period of $5$ s and $0.1$ s, respectively. Solid lines show the corresponding analytical expectation. Since the NS emission is modeled as strictly periodic, the distribution should be interpreted as a rough upper limit on the number of detectable events.}}
    \label{fig:counts}
\end{figure}

\subsection{Black hole mass constraints from waiting times}\label{subsec:MBHSP}

As discussed earlier, the observed activity window of \object{FRB~20121102A} can be reproduced for a wide range of BH masses. Consequently, to break this degeneracy and place meaningful constraints on the BH mass, independent observational indicators are required. One such indicator is the waiting-time distribution between pulses. \citet{cruces} suggested that burst arrival times may exhibit a characteristic timescale distinct from the average Poissonian distribution of regular bursts—a result later supported by \citet{Li2021}, who reported hundreds of bursts separated on average by only a few milliseconds, along with other groups separated by tens of seconds.

{We propose that such clusters of closely spaced millisecond pulses could originate from gravitational lensing, where image splitting produces secondary echoes delayed relative to the primary burst by the combined geometrical and gravitational time delay.} A detailed morphological comparison between lensed and observed FRB bursts is deferred to future work.


{However, lensing-induced image splitting does not imply that both components exceed the detection threshold. The total amplification $\mathcal{A}_{\rm L}$ is shared between the primary image and the echo, with effective echo amplification $\mathcal{A}_{\rm echo}=\mathcal{A}_{\rm L}/(1+R_L)$ (Eq.~\ref{eq:luminosity_ratio}). Near the caustic ($R_L\sim1$), the echo receives approximately half the total flux. For a pulse with the mean intrinsic energy, detecting both components therefore requires $\mathcal{A}_{\rm L}\gtrsim2\mathcal{A}_T$, corresponding to $\log_{10}\mathcal{A}_{\rm L}\gtrsim2.3$ for $\mathcal{A}_T=100$. For individual pulses, the required amplification depends on their intrinsic energy.}

{We adopt an idealized detection prescription based solely on an energy threshold, without modelling an instrument-specific sensitivity or temporal response. After splitting the amplified pulse into primary and secondary components, we apply the energy threshold to each component separately. Both are counted as separate bursts when each exceeds this threshold, irrespective of their temporal separation. Each such pair contributes one count to the primary--echo delay histogram. No minimum time-delay cut is imposed. Finite pulse durations and instrumental temporal resolution could merge some pairs into single observed events, while beaming and propagation effects could further reduce the observable echo fraction; these effects are not included in this detection prescription.}

{This is relevant for systematic searches for lensed FRBs, which have found no statistically significant detections in large samples \citep[e.g.][]{Zhou2022,Kader2022,Leung2022}, though individual candidates have been proposed \citep[e.g.][]{Xiong2026}. Such searches assume two highly correlated burst copies, but in our scenario the two lensed images are emitted along different directions around the Einstein scale; if the NS beam is strongly collimated, as expected for magnetospheric emission \citep[e.g.][]{Voisin,Zhang2023}, it may illuminate only one branch, suppressing the echo entirely. The echo fraction and flux ratio above should thus be treated as upper limits rather than firm predictions—a realistic double-pulse rate requires ray tracing with a collimated, time-dependent beam. Magnetospheric propagation can act as a further selection filter: \citet{Xiao2024} showed that it preferentially suppresses bursts separated by short time lags, reshaping the waiting-time and flux distributions and lowering the observable double-pulse fraction still further.}


Our simulations yield a delay distribution resembling a log-normal shape (red bars in Fig.~\ref{fig:wtimes}).

{Under the interpretation of the millisecond waiting times as lensing-induced image delays, Eq.~(\ref{Eq:image_delay}) can provide an order-of-magnitude estimate of the BH mass, conditional on the orbital geometry and the characteristic projected separation of the observable split bursts. We denote this separation by $b_{\rm peak}$ and parameterize it as $b_{\rm peak}=\lambda r_s$, where $\lambda$ is dimensionless. Its value depends on the orbital sampling, the intrinsic pulse-energy distribution, and the selection of detectable, temporally resolved primary and secondary pulses. In Appendix~\ref{app:lambda}, we explore representative values spanning $0.1\leq\lambda\leq10$ to illustrate how the inferred mass depends on this uncertain separation. This comparison shows that IMBH configurations with $M_{\rm BH}\sim10^3$--$10^4\,M_\odot$ can reproduce the observed millisecond timescale for suitable choices of $\lambda$ and the orbital parameters. This is an illustrative compatible mass range rather than a unique constraint.}

An additional independent lower bound on the BH mass can be inferred from the system’s gravitational wave inspiral time. Since \object{FRB~20121102A} has been active for more than a decade, we require $\tau_{\mathrm{GW}} > 10$~yr. On the other hand, observations show that prFRBs are typically offset from the centres of their host galaxies, implying that their central engines are unlikely to be supermassive BHs, i.e., $M_{\mathrm{BH}} \lesssim 10^6\ M_\odot$.

The second clustering in the waiting-time distribution may emerge from the interplay between the NS’s {cadence} period and its orbital period, as illustrated by the blue bars in Fig.~\ref{fig:wtimes}. The spread in this distribution reflects that not every eclipse produces a detectable pulse: a weak burst may fail to be sufficiently amplified when the NS is behind the BH, or the crossing behind the BH may be faster than the NS spin period. The latter is particularly common at the beginning of the activity window, when the NS is moving near the edge of the BH. Together, these effects generate a harmonic structure in the waiting-time histogram, capturing the combined influence of both {cadence} and orbital periodicities.

{An alternative explanation invokes gravitational self-lensing by the NS itself \citep{DallOsso2024,LaPlaca2026}: one or more magnetospheric hotspots are amplified as they approach the caustic line behind the star, and a two-hotspot configuration naturally yields a weakly lensed population together with a strongly amplified one from the far side of the NS. In that framework, the repetition and waiting-time statistics arise not from a lensing-induced image delay but from the stochastic hotspot activity filtered by rotational/precessional geometry and the detection threshold \citep{DallOsso2024,LaPlaca2026}.}

So far, our model does not reproduce the observed waiting-time peak at $\sim 10^{2}\,\mathrm{s}$; the closest characteristic timescale is the orbital period ($\sim 40$ minutes), which is about an order of magnitude longer. One plausible explanation is emission from a secondary magnetic pole, lensed at a different impact angle, consistent with NICER observations of PSR~J0030+0451 \citep{Bilous2019, Riley2019} and PSR~J0740+6620 \citep{Salmi2022} showing that neutron star magnetic poles are not necessarily antipodal. Other possibilities include an intrinsic emission process operating on $\sim 10^{2}\,\mathrm{s}$ timescales, neutron star precession, or variations in the orbital plane inclination driven by external perturbations.

\begin{figure}
  \centering
   \includegraphics[width=1\linewidth]{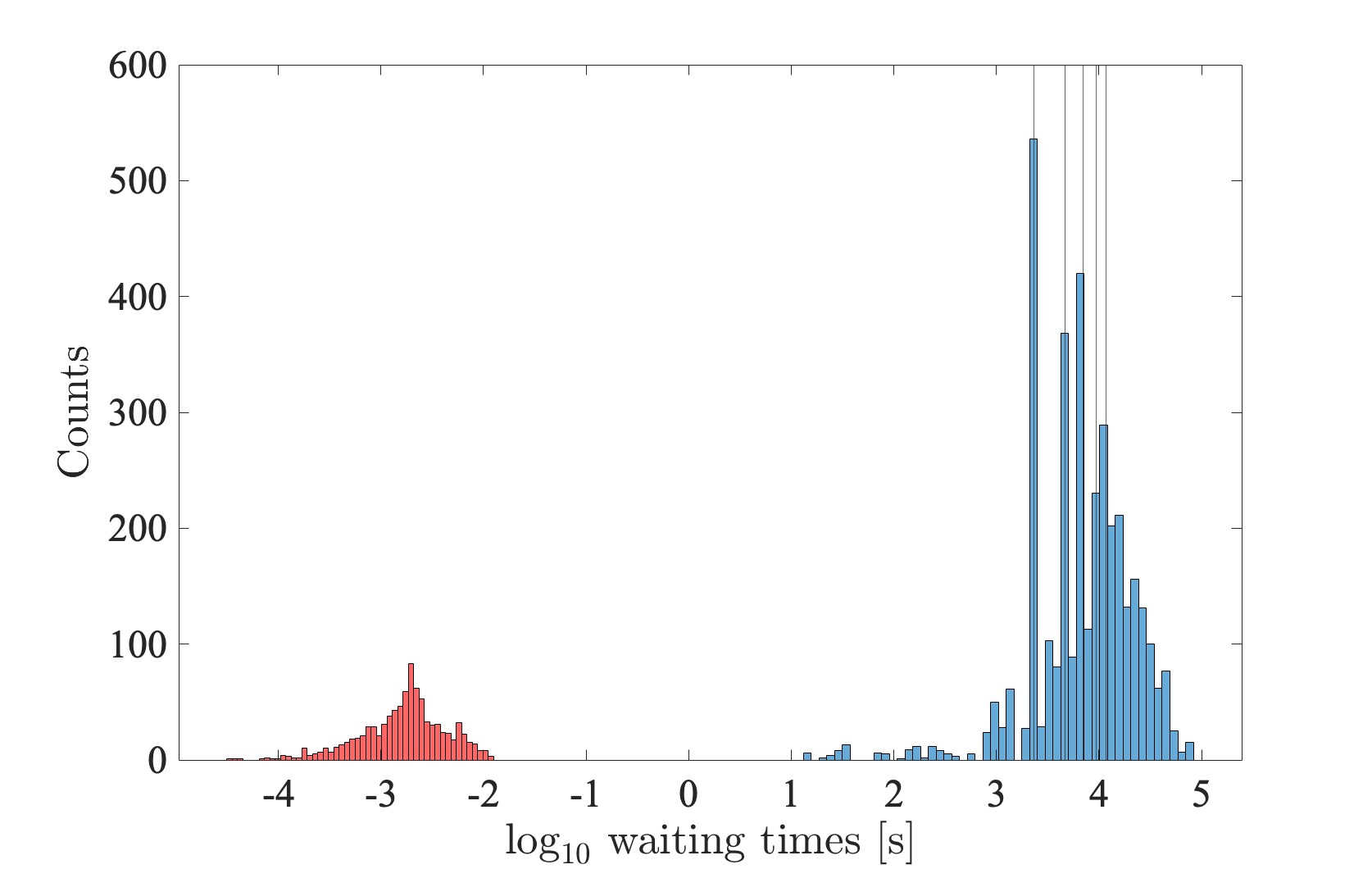}
   \caption{Predicted waiting-time distribution between bursts expected from the LTp model. 
   The blue distribution  corresponds to the orbital motion of the NS around the BH, while the red distribution corresponds to be the {geometrical} delay produced by pulse splitting. The lines over the blue distribution indicate the orbital period and its harmonics. The simulation was performed assuming a BH mass of $M_{\text{BH}} = 10^{3.5}$ M$_{\odot}$, inclination of $i=0.05^{\circ}$ and a NS {cadence} period of $P_{\text{NS}} = 5$ s.   
   }
    \label{fig:wtimes}
\end{figure}

\subsection{Energy distribution}


The energy distribution of lensed pulses originating from an underlying log-normal distribution emitted by a magnetar is shown in Fig.~\ref{fig:lensing_col}. {The primary and secondary images are treated according to the idealized detection prescription described above.} Gravitational lensing transforms the intrinsic log-normal energy distribution of magnetar pulses into a bimodal profile: weakly lensed bursts preserve the original log-normal shape, while strongly lensed bursts populate a high-energy tail that follows a power-law distribution. {For uniform cadence emission and sufficiently dense temporal sampling of the lensing geometry, changing $P_{\rm NS}$ changes the number of pulses sampled over a fixed time interval, but not the expected slope of the energy distribution. It therefore changes the normalization of an unnormalized histogram, corresponding to a vertical shift on logarithmic axes. For sufficiently long emission periods, however, the sampling can miss the brief intervals of strong amplification, reducing the number of pulses above the detection threshold and potentially distorting or suppressing the apparent high-energy tail.}

This bimodality arises naturally within the standard self-lensing framework. Under typical system parameters, however, the expected amplification is insufficient to boost the pulses of a standard Galactic magnetar $10^{29}$--$10^{30}$ erg \citep[e.g.][]{VeraCasanova2025} to the extreme energies observed in \object{FRB~20121102A}, unless the source possesses an intrinsically larger energy budget, with average pulse energies on the order of $\sim10^{34}$ erg. Nonetheless, a magnetar emitting exceptionally strong radio pulses---such as \object{SGR~1935+2154}---could partially overlap the energy range observed for \object{FRB~20121102A}.

Notably, the calculation of the amplification adopted in this work carries uncertainties regarding the region of the neutron star that is effectively magnified. {In our baseline treatment, we approximate the emitting region as a uniformly bright source with the radius of the NS. Emission at other wavelengths could also be lensed, provided that its emitting region has a favourable alignment. This possibility is motivated by the X-ray burst accompanying the FRB-like radio event from SGR~1935+2154. However, the amplification and detectability of such counterparts depend on the size, location, and emission pattern of the corresponding regions, which require a separate investigation.}

Alternatively, {a smaller emitting region, such as a polar cap or a localized region of the magnetosphere, could attain a larger maximum amplification; see Eq.~(\ref{A_max}). However, reaching this enhanced maximum requires a more precise alignment with the caustic, reducing the range of favourable configurations. A smaller source therefore does not automatically imply a larger detectable population. At sufficiently extreme amplifications, wave-optics diffraction must also be considered \citep{Bakala2023}.}

{Furthermore, a fully relativistic treatment would allow us to assess corrections to the weak-deflection approximation $\alpha=2r_s/r_\gamma$. Their importance depends on the photon trajectories as well as the orbital separation, so stronger amplification cannot be inferred solely from the adopted scale $a\sim200r_s$. More compact configurations may exhibit additional strong-field lensing effects \citep{Bakala2023}.}

Taking these uncertainties into account, the overall energy profile may still appear bimodal, with a log-normal component arising from the intrinsic magnetar emission and a high-energy power-law tail potentially produced by gravitational self-lensing. {Lensing-generated high-energy tails have also been obtained for self-lensing by the NS itself \citep{DallOsso2024,LaPlaca2026}, whereas here the lens is the BH companion.} Within this framework, radio pulses from a bright, Galactic-like magnetar could, in principle, be amplified to match the energies observed in \object{FRBs}. {The ray-tracing analysis of \citet{Bakala2023} illustrates amplifications as high as $\mathcal{A}\sim10^8$ for point sources in extreme lensing configurations. These values should not be directly extrapolated to our finite emitting region, whose maximum amplification is limited by its size and, where relevant, wave-optics diffraction.} Therefore, a rigorous assessment of the BH--NS scenario will require detailed ray-tracing simulations that incorporate both the lensing geometry and the effects of beam collimation.

At even smaller orbital separations, additional relativistic effects are expected to emerge: photons may execute multiple orbits around the BH, producing delayed echoes; bursts can be magnified even when the NS is in front of the BH; and the activity window becomes increasingly intricate and time-dependent \citep[e.g.][]{Bakala2023}. Such mechanisms could naturally account for a broader class of repeating FRBs, particularly those lacking strict periodicity but exhibiting similar energy and waiting-time distributions.

\begin{figure}
    \centering
    \includegraphics[width=1\linewidth]{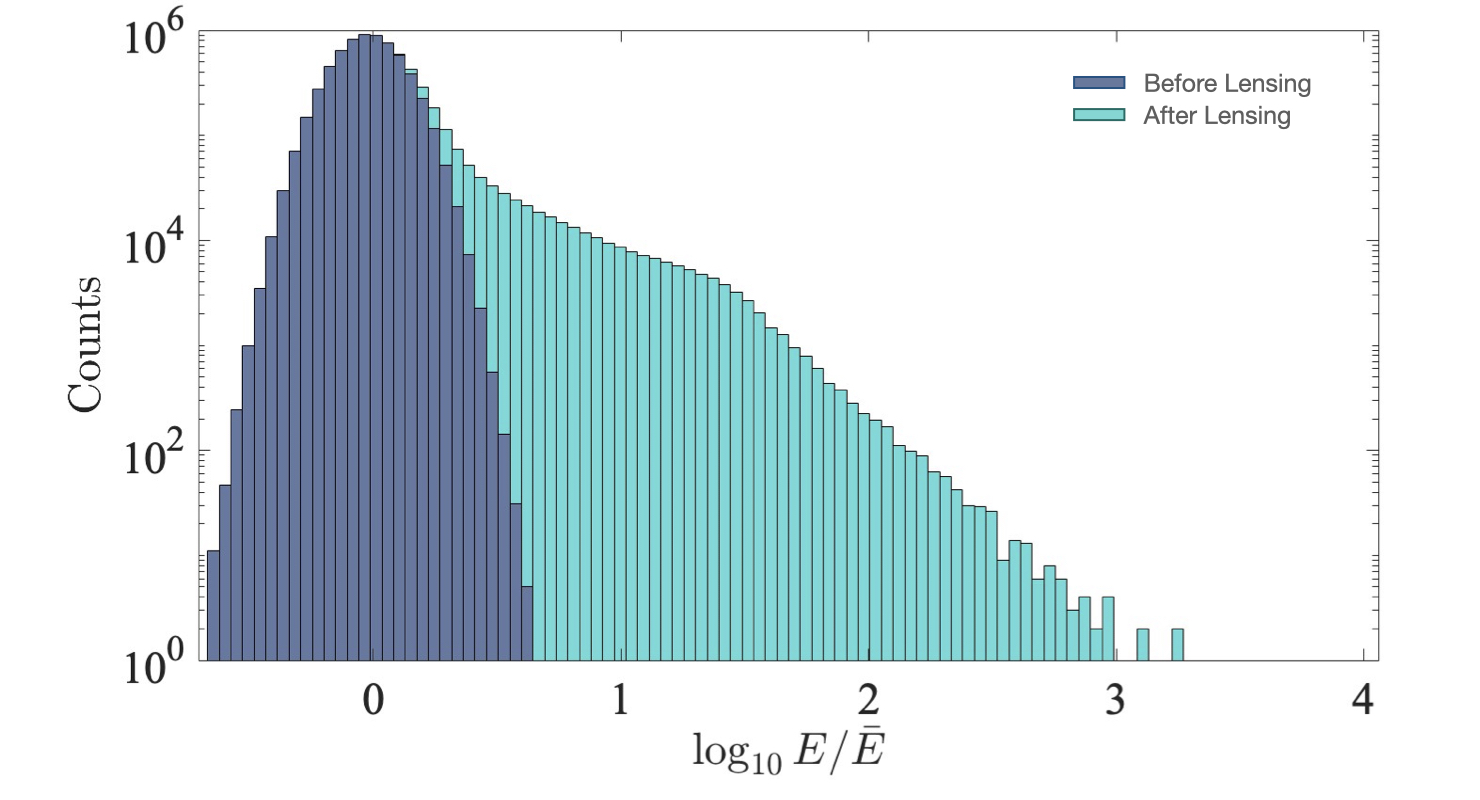}
    \caption{Energy distribution of radio pulses before (purple bars) and after lensing (light turquoise bars). Energies are normalized to the average intrinsic emission of the magnetar. In this example, we adopt $\sigma = 0.1$ for the underlying log-normal distribution. For the simulation we consider a BH mass $10^{3.5}$ M$_{\odot}$, inclination angle $i=0.05^{\circ}$ and {magnetar} {cadence} period $P=5$ s.}
    \label{fig:lensing_col}
\end{figure}

\subsection{{Expected number of observable IMBH--NS systems}}\label{sec:Number_of_systems}

{A central question for the viability of the present scenario is whether enough IMBH--NS systems could exist to account for the small number of periodically repeating FRBs currently known. We emphasize that our model is not intended to explain the full FRB population, but rather a rare sub-channel associated with periodic repeaters. We therefore estimate the number of potentially observable systems only at the order-of-magnitude level. A detailed discussion of the selection factors entering this estimate is given in Appendix~\ref{app:population}.}

{Using the factorized population estimate described there, we obtain}
\begin{equation}
N_{\rm obs}
\lesssim
{1}
\left(\frac{N_{\rm IMBH}}{10^{8}}\right)
\left(\frac{f_{\rm off/SF}}{0.1}\right)
\left(\frac{f_{\rm yMag/tight}}{10^{-3}}\right)
\left(\frac{f_{\rm radio}}{0.1}\right)
\left(\frac{f_{\rm geom}}{10^{-3}}\right).
\label{eq:Nobs_final}
\end{equation}
{Here, $N_{\rm IMBH}$ is the parent population of IMBHs in the adopted mass and redshift ranges; $f_{\rm off/SF}$ is the fraction residing in suitable off-nuclear or star-forming environments; $f_{\rm yMag/tight}$ is the fraction of this population hosting an active young magnetar in a sufficiently compact, surviving orbit; $f_{\rm radio}$ is the radio-emitting fraction of magnetars; and $f_{\rm geom}$ is the fraction satisfying the required viewing geometry. The adopted values and their uncertainties are discussed in Appendix~\ref{app:population}.}

{For the adopted fiducial values, this estimate gives
$N_{\rm obs}\sim1$, suggesting that the proposed channel would
be very rare under these assumptions. This normalization is subject
to substantial uncertainties in the parent IMBH abundance,
the fraction residing in suitable environments, the occurrence
of compact IMBH--young-magnetar systems, the radio-emitting
fraction of magnetars, and the effective geometrical selection.
Because these factors enter multiplicatively, lower values
for several of them could reduce the expected number by orders
of magnitude. Additional observational selection effects would
further reduce the number of detected systems. The estimate
should therefore be regarded as exploratory rather than as
a robust prediction of the observable population.}

\section{Conclusions}\label{sec:conclusions}

{We explored whether radio pulses emitted by a NS orbiting a BH could be amplified through self-lensing and account for periodically repeating FRBs. Purely Keplerian orbits fail to reproduce the observed activity windows, producing windows lasting only minutes. In contrast, inclined orbits undergoing Lense--Thirring precession can produce activity windows spanning a significant fraction of the FRB cycle. The duration of the active phase depends mainly on the BH mass, the orbital inclination, and the amplification threshold, while its exact boundaries remain diffuse because they also depend on the intrinsic pulse-energy distribution and telescope sensitivity.}

{The millisecond component of the waiting-time distribution can be naturally associated with lensing-induced image delays between a primary burst and its secondary echo. For \object{FRB~20121102A}, interpreting the shortest waiting times in this way favours an intermediate-mass BH with $M_{\rm BH}\sim10^3$--$10^4 M_\odot$. By contrast, the longer $\sim10^2$ s waiting-time component is unlikely to be produced solely by the orbital period, which is typically closer to $\sim10^3$ s in our fiducial models. A possible explanation is that emission from a secondary, non-antipodal magnetic pole is also lensed along a different trajectory, producing additional delays of order $10^1$--$10^2$ s.}

{The lensed energy distribution can develop a high-energy tail from an intrinsic log-normal distribution of magnetar-like radio pulses. However, within the isotropic weak-lensing treatment adopted here, the amplification is not sufficient to bridge the full energy gap between Galactic magnetars and the brightest FRBs. The scenario therefore requires either an intrinsically energetic, recurrent magnetar-like source, stronger amplification from smaller emitting regions, beaming effects, or strong-lensing contributions.}

{The model is also subject to strong population and geometric selection effects. Our order-of-magnitude estimate suggests that IMBH--NS self-lensing systems could plausibly account for a very small subpopulation of periodically repeating FRBs, but not for the full FRB population. The dominant uncertainties are the formation efficiency of compact IMBH--young-magnetar binaries, the fraction of rapidly spinning IMBHs, and the probability of the required line-of-sight alignment.}

{A proper assessment of this scenario will require relativistic ray-tracing simulations that include the detailed lensing geometry, beam collimation, magnetospheric propagation, and the finite size and location of the emitting region.}

\begin{acknowledgements}
This work is funded by the Max Planck Society through the Max Planck Partner Group led by Prof. M. Cruces. M. Cruces acknowledges funds from grant ANID CIA250016 (CPS-RTC). {We thank the anonymous referee for their valuable comments and suggestions, which helped to improve the clarity and scope of this manuscript. We also thank Juliette Bouclet for carefully reviewing  the equations used in this work.}      
\end{acknowledgements}

%

\begin{thebibliography}{}

\bibitem[Anna-Thomas et al.(2023)]{AnnaThomas2023}
Anna-Thomas, R., Connor, L., Dai, S., et al. 2023, Science, 380, 599.

\bibitem[Akg{\"u}n et al.(2006)]{Akgun2006} Akg{\"u}n, T., Link, B., Wasserman, I. 2006, MNRAS, 365, 653

\bibitem[Bakala et al.(2023)]{Bakala2023}
Bakala, P., Bakalová, A., La Placa, R., Falanga, M., \& Stella, L. 2023, A\&A, 673, A164.


\bibitem[Backer \& Hellings(1986)]{Backer1986} Backer, D.~C. \& Hellings, R.~W. 1986, ARA\&A, 24, 537


\bibitem[Bardeen et al.(1972)]{Bardeen1972}
Bardeen, J. M., Press, W. H., \& Teukolsky, S. A. 1972, ApJ, 178, 347.

\bibitem[Berti \& Volonteri(2008)]{Berti2008}
Berti, E., \& Volonteri, M. 2008, ApJ, 684, 822.

\bibitem[Beckmann et al.(2023)]{Beckmann2023}
Beckmann, R. S., Dubois, Y., Volonteri, M., et al. 2023, MNRAS, 523, 5610

\bibitem[Beloborodov(2021)]{Beloborodov2021}
Beloborodov, A. M. 2021, ApJL, 922, L7



\bibitem[Bilous et al.(2019)]{Bilous2019} 
Bilous, A. V., Riley, T. E., Watts, A. L., et al. 2019, ApJL, 887, L23.


\bibitem[Blandford \& Teukolsky(1976)]{Blandford1976} Blandford, R. \& Teukolsky, S.~A. 1976, ApJ, 205, 580


\bibitem[Braga et al.(2025)]{Braga2025} Braga, C.~A., Cruces, M., Cassanelli, T., et al. 2025, A\&A, 693, A40

\bibitem[Burke-Spolaor et al.(2012)]{Burke2012} Burke-Spolaor, S., Johnston, S., Bailes, M., et al. 2012, MNRAS, 423, 1351

\bibitem[CHIME/FRB Collaboration et al.(2020a)]{CHIME2020periodic}
CHIME/FRB Collaboration, Amiri, M., Andersen, B.~C., et al. 2020a, Nature, 582, 351

\bibitem[CHIME/FRB Collaboration et al.(2020b)]{Andersen2020}
CHIME/FRB Collaboration, Andersen, B.~C., Bandura, K.~M., et al. 2020b, Nature, 587, 54

\bibitem[Cruces et al.(2021)]{cruces} Cruces, M., Spitler, L.~G., Scholz, P., et al. 2021, MNRAS, 500, 448

\bibitem[Dall'Osso et al.(2024)]{DallOsso2024}
Dall'Osso, S., La Placa, R., Stella, L., Bakala, P., \& Possenti, A. 2024, ApJ, 973, 123

\bibitem[Desvignes et al.(2018)]{Desvignes2018}
Desvignes, G., Eatough, R. P., Pen, U. L., et al. 2018, ApJL, 852, L12

\bibitem[Desvignes et al.(2024)]{Desvignes2024}
Desvignes, G., Weltevrede, P., Gao, Y., et al. 2024,
Nature Astronomy, 8, 617--627.

\bibitem[Di Carlo et al.(2021)]{DiCarlo2021}
Di Carlo, U. N., Mapelli, M., Pasquato, M., et al.
2021, MNRAS, 507, 5132.

\bibitem[Eftekhari et al.(2024)]{Eftekhari2024}
Eftekhari, T., Dong, Y., Fong, W., Shah, V., Simha, S., Andersen, B. C., Andrew, S., Bhardwaj, M., \textit{et al.} 2024, ApJL, 979, L21

\bibitem[Espinoza-Dupouy et al.(2025)]{EspinozaDupouy2025}
Espinoza-Dupouy, M.~C., Cruces, M., Cassanelli, T., Braga, C.~A., Bermúdez, E., \& Vera-Casanova, J. 2025, arXiv:2507.04609

\bibitem[Gajjar et al.(2018)]{Gajjar2018} 
Gajjar, V., Siemion, A.~P.~V., Price, D.~C., et al.\ 2018, ApJ, 863, 2. doi:10.3847/1538-4357/aad005


\bibitem[Feng et al.(2024)]{Feng2024} 
Feng, X.-M., Yang, Y.-P., \& Li, Q.-C. 2024, MNRAS, 530, 3641.  

\bibitem[Gerosa et al.(2021)]{Gerosa2021}
Gerosa, D., Giacobbo, N., \& Vecchio, A. 2021, ApJ, 915, 56.

\bibitem[Greene et al.(2020)]{Greene2020}
Greene, J. E., Strader, J., \& Ho, L. C. 2020, ARA\&A, 58, 257

\bibitem[Hessels et al.(2019)]{Hessels2019}
Hessels, J. W. T., Spitler, L. G., Seymour, A. D., et al. 2019, ApJL, 876, L23

\bibitem[Hilmarsson et al.(2021)]{Hilmarsson2021}
Hilmarsson, G.~H., Michilli, D., Spitler, L.~G., et al. 2021, ApJ, 908, L10

\bibitem[Hilmarsson et al.(2021)]{Hilmarsson2021b}
Hilmarsson, G.~H., Spitler, L.~G., Main, R.~A., \& Li, D.~Z. 2021, MNRAS, 508, 5354

\bibitem[Ioka \& Zhang (2020)]{Ioka}
Ioka, K., \& Zhang, B. 2020, ApJL, 893, L26

\bibitem[Josephy et al.(2019)]{Josephy2019}
Josephy, A., Chawla, P., Fonseca, E., et al. 2019, ApJL, 882, L18

\bibitem[Kader et al.(2022)]{Kader2022}
Kader, Z., Leung, C., Dobbs, M., et al. 2022, Phys. Rev. D, 106, 043016


\bibitem[Kritos et al.(2024)]{Kritos2024}
Kritos, K., Reali, L., Gerosa, D., \& Berti, E. 2024,
Phys. Rev. D, 110, 123017,
doi:10.1103/PhysRevD.110.123017.

\bibitem[Kumar et al.(2021)]{Kumar2021}
Kumar, P., Shannon, R. M., Flynn, C., et al. 2021, MNRAS, 500, 2525.


\bibitem[La Placa et al.(2026)]{LaPlaca2026}
La Placa, R., Dall'Osso, S., Stella, L., \& Possenti, A. 2026, arXiv e-prints, arXiv:2603.12386

\bibitem[Leung et al.(2022)]{Leung2022}
Leung, C., Kader, Z., Masui, K. W., et al. 2022, Phys. Rev. D, 106, 043017

\bibitem[Levin et al.(2020)]{Levin2020}
Levin, Y., Beloborodov, A. M., \& Bransgrove, A. 2020, ApJL, 895, L30


\bibitem[Li et al.(2022)]{Li2022}
Li, Q.-C., Yang, Y.-P., Wang, F.~Y., Xu, K., \& Dai, Z.-G. 2022, MNRAS, 517, 4612

\bibitem[Li et al.(2021)]{Li2021} Li, D., Wang, P., Zhu, W.~W., et al. 2021, Nature, 598, 267

\bibitem[Lorimer et al.(2007)]{Lorimer} Lorimer, D.~R., Bailes, M., McLaughlin, M.~A., et al. 2007, Science, 318, 777

\bibitem[Liebes(1964)]{Liebes1964} Liebes, S. 1964, Phys. Rev., 133, B835

\bibitem[Lyutikov et al.(2020)]{Lyutikov} Lyutikov, M., Barkov, M.~V., \& Giannios, D. 2020, ApJ, 893, L39

\bibitem[Lyutikov(2024)]{Lyutikov2024}
Lyutikov, M. 2024, MNRAS, 529, 2180

\bibitem[Maeder(1973)]{Maeder1973} Maeder, A. 1973, A\&A, 26, 215  

\bibitem[Merritt(2013)]{Merritt2013} Merritt, D. 2013, Dynamics and Evolution of Galactic Nuclei (Princeton: Princeton Univ. Press)

\bibitem[Pal(2025)]{Pal2025} 
Pal, A. 2025, arXiv e-prints, arXiv:2502.11215

\bibitem[Patruno et al.(2005)]{Patruno2005}
Patruno, A., Colpi, M., Faulkner, A., \& Possenti, A.
2005, MNRAS, 364, 344.


\bibitem[Pastor-Marazuela et al.(2021)]{Pastor2021} Pastor-Marazuela, I., Connor, L., van Leeuwen, J., et al. 2021, Nature, 596, 505

\bibitem[Rajwade et al.(2020)]{Rajwade} Rajwade, K.~M., Mickaliger, M.~B., Stappers, B.~W., et al. 2020, MNRAS, 495, 3551

\bibitem[Rea \& De Grandis(2025)]{Rea2025} 
Rea, N. \& De Grandis, D. 2025, arXiv e-prints, arXiv:2503.04442

\bibitem[Reynolds(2021)]{Reynolds2021}
Reynolds, C. S. 2021, ARA\&A, 59, 117.


\bibitem[Riley et al.(2019)]{Riley2019} 
Riley, T. E., Watts, A. L., Bogdanov, S., et al. 2019, ApJL, 887, L21.


\bibitem[Salmi et al.(2022)]{Salmi2022} 
Salmi, T., Riley, T. E., Degenaar, N., et al. 2022, ApJ, 941, 150.

\bibitem[Schneider(1992)]{schneider}
Schneider, P., Ehlers, J., \& Falco, E. E. 1992, Gravitational Lenses (Springer-Verlag, Berlin)

\bibitem[Shaham(1977)]{Shaham1977} 
Shaham, J. 1977, ApJ, 214, 251.  

\bibitem[Spitler et al.(2014)]{Spitler} Spitler, L.~G., Cordes, J.~M., Hessels, J.~W.~T., et al. 2014, ApJ, 790, 101

\bibitem[Spitler et al.(2016)]{Spitler2016}
Spitler, L. G., Scholz, P., Hessels, J. W. T., et al. 2016, Nature, 531, 202.

\bibitem[Swartz et al.(2011)]{Swartz2011}
Swartz, D. A., Soria, R., Tennant, A. F., \& Yukita, M.
2011, ApJ, 741, 49.

\bibitem[Tendulkar et al.(2017)]{Tendulkar2017}
Tendulkar, S. P., Bassa, C. G., Cordes, J. M., et al. 2017, ApJ Lett., 834, L7.

\bibitem[Vera-Casanova et al.(2025)]{VeraCasanova2025}
Vera-Casanova, J., Cruces, M., Liu, K., Wongphexhauxsorn, J., Braga, C. A., Kramer, M., Torne, P., Limaye, P., Espinoza-Dupouy, M. C., Rodríguez, L., et al. 2025, A\&A, aa55088-25.

\bibitem[Voisin(2023)]{Voisin} Voisin, G. 2023, A\&A, 675, A200

\bibitem[Volonteri et al.(2005)]{Volonteri2005}
Volonteri, M., Madau, P., Quataert, E., \& Rees, M. J. 2005, ApJ, 620, 69.

\bibitem[Wada et al.(2021)]{Wada2021} 
Wada, T., Ioka, K., \& Zhang, B. 2021, ApJ, 920, 54.  


\bibitem[Xiao et al.(2024)]{Xiao2024}
Xiao, D., Dai, Z.-G., \& Wu, X.-F. 2024, ApJ, 962, 35

\bibitem[Xiong et al.(2026)]{Xiong2026}
Xiong, S.-W., Xiao, S., Jiang, Z.-H., Lai, Y., Zhang, Y.-Q., Zhao, R.-S., \& You, Z.-Y. 2026, ApJ, 996, 29

\bibitem[Xu et al. (2022)]{Xu2022}
Xu, H., Niu, J. R., Chen, P., et al. 2022, Nature, 609, 685,

\bibitem[Zanazzi \& Lai(2020)]{Zanazzi2020}
Zanazzi, J. J., \& Lai, D. 2020, ApJL, 892, L15

\bibitem[Zhang(2020)]{Zhang2020} Zhang, B. 2020, Nature, 587, 45

\bibitem[Zhang et al. (2021)]{Zhang2021} Zhang, G. Q., et al. 2021, ApJL, 920, L23

\bibitem[Zhang et al. (2023)]{Zhang2023a} Zhang, Y.-K., Li, D., Zhang, B., et al. 2023, ApJ, 955, 142
 
\bibitem[Zhang(2023)]{Zhang2023} Zhang, B. 2023, Rev. Mod. Phys., 95, 035005


\bibitem[Zhou et al.(2022)]{Zhou2022}
Zhou, H., Li, Z., Liao, K., Niu, C., Gao, H., Huang, Z., Huang, L., \& Zhang, B. 2022, ApJ, 928, 124







\end{thebibliography}
%

\appendix

\section{Orbit construction}
\label{app:orbit}

Here we give the explicit equations used to construct the NS orbit, summarized only schematically in Sect.~\ref{subsec:equations}.

For the geometry of the orbit, we consider that LTp acts directly on a Keplerian-like orbit, where the radial distance $r$ of the NS to the BH is given by
\begin{equation}\label{Kepler}
r(\nu)= \frac{a(1-e^2)}{1+e\cos(\nu)},
\end{equation}
where the semi-major axis $a$ is inferred from Eq.~(\ref{T_LT}), while the true anomaly $\nu$ is computed from the orbital period as
\begin{equation}\label{tanomaly}
\nu = \frac{2\pi}{\tau_{\rm{Orb}}}t.
\end{equation}

The secular precession rate of the argument of periapsis, including both the Schwarzschild contribution and the leading spin--orbit contribution associated with the BH angular momentum, is \citep[e.g.][]{Merritt2013}{
\begin{equation}\label{Einstein_P}
   \biggl \langle \frac{d\omega}{dt} \biggr \rangle
   =
   \frac{6\pi G M_{\rm BH}}{c^2 a(1-e^2)\tau_{\rm Orb}}
   -
   \frac{6G^2M_{\rm BH}^2\chi}{c^3 a^3 (1-e^2)^{3/2}}\cos i .
\end{equation}}
where $\omega$ is the argument of periapsis and $i$ is the orbital inclination. The first term corresponds to the Schwarzschild correction, while the second term represents the Lense--Thirring contribution.

To build the orbit of the NS around the BH, we start from the Keplerian ellipse, whose rectangular coordinates $x_{NS}$ and $y_{NS}$ in a non-rotating frame are
\begin{equation} \label{rec_coords}
\begin{split}
x_{NS} & = r(\nu )\cos(\nu ), \\
y_{NS} & = r(\nu )\sin(\nu ).
\end{split}
\end{equation}
The periapsis of this orbit then rotates the frame defined in Eq.~(\ref{rec_coords}) following the precession from Eq.~(\ref{Einstein_P}),
\begin{equation} \label{rot_cord}
\begin{split}
x_{NSr}&= x_{NS}\cos(\omega) -y_{NS}\sin(\omega ), \\
y_{NSr}&= x_{NS}\sin(\omega)+y_{NS}\cos(\omega),
\end{split}
\end{equation}
where $x_{NSr}$ and $y_{NSr}$ are the coordinates of the orbit including the precession of periapsis. Finally, the periapsis-precessing orbit is tilted by an angle $i$ with respect to the BH equatorial plane, defined perpendicular to the BH spin axis, and its line of nodes precesses on the Lense--Thirring timescale given by Eq.~(\ref{T_LT}),
\begin{equation} \label{Obs_coords}
\begin{split}
x_{Obs}& = x_{NSr}\cos(\Omega) - y_{NSr}\sin(\Omega) \cos(i),\\
y_{Obs}&= x_{NSr}\sin(\Omega) + y_{NSr}\cos(\Omega) \cos(i),\\
z_{Obs}&= y_{NSr}\sin(i).
\end{split}
\end{equation}
The coordinates $x_{Obs}, y_{Obs}, z_{Obs}$ represent the orbit of the NS as seen by a distant observer, combining the Keplerian motion, the precession of periapsis, the Lense--Thirring nodal precession, and the orbital inclination.

\section{Relativistic corrections and Kerr metric}
\label{app:relativistic_corrections}

{In relativistic ray-tracing treatments of compact-object lensing, the observed flux is commonly corrected by a redshift factor $g^4$ \citep[e.g.][]{Bakala2023}. We do not include this factor in Eq.~(\ref{EAmp}) because, in the parameter range explored here, the emitter is located far from the BH horizon. For the fiducial LTp configuration, $a/r_s\sim200$, the gravitational contribution is $g_{\rm grav}^4\simeq(1-r_s/a)^2\simeq0.99$. The orbital velocity is also only mildly relativistic, $v/c\simeq(r_s/2a)^{1/2}\simeq0.05$, so Doppler corrections are expected to be at most of order unity. Therefore, the omission of $g^4$ may introduce small phase-dependent flux corrections, but it cannot change the amplification by the several orders of magnitude relevant to the energy-gap problem.}

{We also neglect spin-dependent corrections to the photon deflection. At leading order, the Schwarzschild deflection is $\alpha_{\rm Schw}\simeq2r_s/R$, while the Kerr frame-dragging correction scales as $\alpha_{\rm Kerr}\sim\chi r_s^2/r_{\gamma}^2$, where $R$ is the photon impact parameter around the BH. Their ratio is therefore $\alpha_{\rm Kerr}/\alpha_{\rm Schw}\sim\chi r_s/(2r_{\gamma})$. For the fiducial LTp configuration, $a/r_s\sim200$, the relevant Einstein scale is $R_E\simeq\sqrt{2r_sa}\simeq20r_s$, giving $\alpha_{\rm Kerr}/\alpha_{\rm Schw}\sim\chi/40\lesssim3\%$ even for $\chi\simeq1$. Thus, at the Einstein-ring scale that dominates the weak-deflection amplification, the spin correction to the bending angle is much smaller than the Schwarzschild term.}

\section{Timing corrections}
\label{app:timing_corrections}

{In addition to the lensing-induced image delay between the primary and echo components, the observed arrival time of each pulse can be affected by standard timing corrections in compact binaries. In this work, we include the Rømer delay, the Einstein delay, and the orbital Shapiro delay. These terms affect the arrival time of consecutive pulses emitted at different orbital phases, and should therefore be distinguished from the image delay produced by the splitting of a single burst.}

{The Rømer delay accounts for the geometrical light-travel-time difference caused by the changing position of the NS along its orbit \citep{Blandford1976}. It is given by}
\begin{equation}
\tau_{\rm R}= (1+z)\frac{\vec{r}_{\mathrm{Obs}}\cdot \vec{n}_{\odot}}{c},
\end{equation}
{where $\vec{r}_{\mathrm{Obs}}$ is the position vector of the NS in the observer frame, $\vec{n}_{\odot}$ is the unit vector pointing from the Solar System barycenter to the binary-system barycenter, and $z$ is the redshift of the host galaxy.}

{The Einstein delay combines gravitational redshift and special-relativistic time dilation effects. At first post-Newtonian order, we write}
\begin{equation}
\tau_{\rm E}=(1+z)\frac{2GM_{\rm BH}}{c^3}\ln\left[1-e\cos(\nu)\right],
\end{equation}
{where $e$ is the orbital eccentricity and $\nu$ is the true anomaly. This contribution vanishes for the circular fiducial orbits used in the main simulations.}

{Finally, the orbital Shapiro delay describes the additional light-travel time introduced by the curved spacetime near the BH as the NS moves along its orbit. Following \citet{Backer1986}, we use}

\begin{equation}
\tau_{\mathrm{SD}}= -(1+z) \frac{2GM_{\mathrm{BH}}}{c^3}
\ln \left( |\vec{r}_{\mathrm{Obs}} - \vec{r}_{\mathrm{NS}}|
+ (\vec{r}_{\mathrm{Obs}} - \vec{r}_{\mathrm{NS}}) \cdot \vec{n}_\odot \right),
\label{eq:shapiro_trad}
\end{equation}
{where the BH is located at the origin, $\vec{r}_{\mathrm{NS}}$ is the NS position, and $\vec{r}_{\mathrm{Obs}}$ denotes the observer position.}

\section{Robustness of the BH-mass estimate to the characteristic lensing impact parameter}
\label{app:lambda}

In Sect.~\ref{subsec:MBHSP}, we use the millisecond component of the waiting-time distribution as a possible signature of gravitationally lensed split bursts. In this interpretation, the relevant timescale is the lensing-induced image delay between the primary burst and its secondary echo. This delay depends not only on the BH mass, but also on the projected source--lens separation at which the resolved split bursts are preferentially produced.

If we estimate the projected distance as \( b = r_s / 2 \) to obtain an order-of-magnitude estimation of Eq. (\ref{Eq:image_delay}), this gives \( \bar{b} \sim \mathcal{O}(0.1) \). In this regime, the root squared term in Eq.~\eqref{tau} dominates over the logarithmic contribution.
Therefore, at the relevant scales of mass and distance in this study, Eq. (\ref{tau})  {gives a value} of order  \( \tau \sim \mathcal{O}(0.1) \), providing an estimation of the delay between the main and echo pulses in Eq. (\ref{Eq:image_delay}) of {($\tau_{\mathrm{lens}}^{\mathrm{Spl}} \sim \mathcal{O}(10^{-2}-10^{-3} \mathrm{s}\,)$}.

Expanding Eq. (\ref{Eq:image_delay}) to first order yields an approximate expression for the {delay} between a pulse and its gravitationally lensed echo:{
\begin{equation}\label{Shap}
\tau_{\mathrm{lens}}^{\mathrm{Spl}}\sim  \frac{4GM_{\mathrm{BH}}}{c^3}(1 + z)\bar{b}.
\end{equation}}
This expression reaches its minimum at full eclipse (\( \bar{b} = 0 \)). This is because, at the caustic line, both light paths converge to form an Einstein ring, leading to the simultaneous arrival of the direct pulse and its lensed echo. 

To explore Eq. (\ref{Shap})  we denote the characteristic projected separation by $b_{\rm peak}$. This quantity should not be interpreted as the point of maximum amplification, which occurs at exact alignment, $b=0$. Instead, $b_{\rm peak}$ is an effective distance that indicates where the distribution of observable split bursts is expected to peak. Its value depends on the orbital geometry, the intrinsic pulse-energy distribution, the detection threshold, and the flux sharing between the primary and secondary lensed images. Since these effects are not modelled in detail here, we parameterize this uncertainty as $b_{\rm peak} = \lambda r_s,$ where $\lambda$ is a dimensionless parameter. We consider representative values in the range {$0.1\lesssim\lambda\lesssim10$}, which cover cases where the resolved split bursts are dominated either by events very close to the caustic line or by intrinsically brighter pulses that remain detectable at larger projected separations.
Using Eq.~(\ref{Shap}) evaluated at $b=b_{\rm peak}$, the characteristic image delay can be written as

\begin{equation}
    \tau_{\rm lens}^{\rm Spl}(b_{\rm peak})
    =
    \frac{4GM_{\rm BH}}{c^3}(1+z)\,
    \tau\left(
    \frac{\lambda r_s}{\sqrt{2r_s a}}
    \right),
\end{equation}
where $\tau(\bar b)$ is the dimensionless lensing-delay function and $R_E\simeq \sqrt{2r_s a}$ is the Einstein-ring radius in the source plane. For the small projected separations relevant to the most strongly amplified bursts, $\bar b\lesssim1$, this expression can be approximated, up to factors of order unity, as
\begin{equation}
    \tau_{\rm lens}^{\rm Spl}
    \sim
    \frac{4GM_{\rm BH}}{c^3}(1+z)\,
    \lambda
    \left(\frac{r_s}{a}\right)^{1/2}.
    \label{eq:delay_lambda}
\end{equation}

Figure~\ref{fig:Shapiro_constrain} shows the relation between $M_{\rm BH}$ and $\tau_{\rm lens}^{\rm Spl}$ for several representative values of $\lambda$. The grey band marks the observed millisecond waiting-time component of FRB~20121102A, approximately $10^{-3}$--$10^{-2}$~s. {The values $\lambda=0.1$, $1$, and $10$ explore characteristic projected separations spanning two orders of magnitude, while $\lambda=0.2$ illustrates the sensitivity to a smaller change at the lower end. These choices provide a sensitivity analysis rather than an independently constrained range for $\lambda$. The intersection of each curve with the observed band identifies the compatible BH masses for that assumed characteristic separation and the adopted orbital parameters.}

\begin{figure}
    \centering
    \includegraphics[width=1\linewidth]{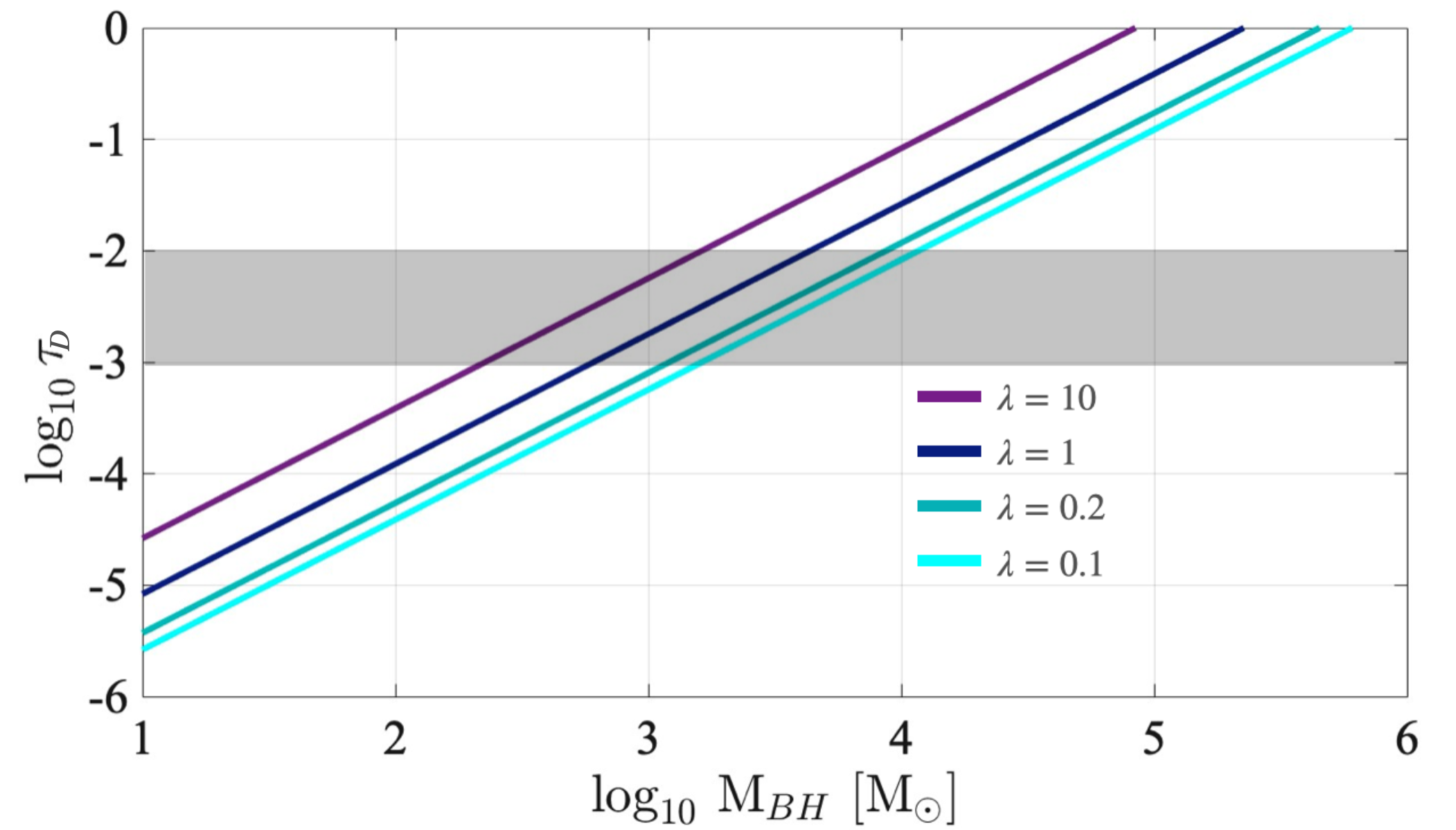}
    \caption{Theoretical relation between BH mass and the lensing-induced delay observed between main and echo pulses after gravitational lensing. The grey horizontal band indicates the observed waiting times for \object{FRB 20121102A}, {which range approximately from $10^{-3}$--$10^{-2}\,{\rm s}$. The lines represent different scenarios in which the peak of the amplified-burst distribution occurs at projected distance} $b=\lambda r_s$, see Eq. (\ref{eq:delay_lambda}).} \label{fig:Shapiro_constrain}
\end{figure}

\section{Population estimate for IMBH--NS systems}
\label{app:population}

{Here we summarize the order-of-magnitude population estimate used in Sect.~\ref{sec:Number_of_systems}. We write the number of potentially observable sources as}
\begin{equation}
N_{\rm obs}\sim
N_{\rm IMBH}\,
f_{\rm off/SF}\,
f_{\rm yMag/tight}\,
f_{\rm radio}\,
f_{\rm geom}.
\end{equation}
{where $N_{\rm IMBH}$ is the number of IMBHs in the relevant mass range, $f_{\rm off/SF}$ is the fraction of IMBHs located in environments compatible with recent star formation or off-nuclear positions, $f_{\rm yMag/tight}$ is the fraction of IMBHs hosting a young magnetar or NS in a sufficiently compact orbit, $f_{\rm radio}$ is the fraction of magnetars capable of producing radio emission and $f_{\rm geom}$ is the geometric alignment probability.}


The abundance of IMBHs is uncertain, especially at the low-mass end. Nevertheless, cosmological simulations and observationally motivated black-hole mass functions suggest that dwarf galaxies may host a large population of low-mass black holes \citep[e.g.][]{Beckmann2023,Greene2020}. {For the mass range $M_{\rm BH}\sim10^3$--$10^5 M_\odot$, we assume a constant comoving number density and scaling by comoving volume gives $N_{\rm IMBH}(<0.2)\sim1.5\times10^7$ and $N_{\rm IMBH}(<0.5)\sim1.9\times10^8$. The first limit encompasses the three prFRBs considered here, while the second defines a broader reference volume motivated by the detection of the repeating source FRB~20190711A at $z=0.522$ \citep{Kumar2021}. This detection establishes that repeating FRBs can be observed at such distances, but does not imply completeness for the population considered in our model. Detectability depends on intrinsic pulse energies, lensing amplification, and survey sensitivity, while identifying periodic activity additionally requires sufficient temporal coverage. The above IMBH counts therefore represent uncertain parent-population normalizations that scale directly with the assumed abundance. Since the observable-population estimate below does not explicitly include these additional selection effects, it should be interpreted as an optimistic count under the adopted assumptions rather than a prediction of survey detections.}

{We adopt $f_{\rm off/SF}\sim0.1$ as an order-of-magnitude
environmental selection factor, motivated by the off-centre
BH population predicted in dwarf galaxies by
\citet{Beckmann2023} and by the location of FRB~20121102A
away from the optical centre of its star-forming dwarf host.} {The observed magnetar population provides an empirical reference for $f_{\rm radio}$: radio emission has been detected from six of the 24 confirmed magnetars considered by \citet{Desvignes2024}, corresponding to a fraction of order $10^{-1}$. } 


{Our fiducial choice $\chi=0.9$ illustrates a configuration
with strong Lense--Thirring precession, but such a high spin
is not required by the model. As shown in
Table~\ref{tab:spin_dependence}, configurations with
$0.1\leq\chi\leq0.9$ preserve the adopted activity cycle and,
within the analytical approximation, a 50\% activity fraction
by adjusting the orbital separation and inclination.
Over this range, the maximum amplification remains of the
same order of magnitude, while the gravitational-wave
coalescence time exceeds the observed activity baseline.}

{The expected spin also depends on the BH growth history.
Successive mergers with lower-mass companions and randomly
oriented angular momenta tend to reduce the dimensionless
spin as the IMBH grows \citep{Gerosa2021,Kritos2024}. However, if the
companion mass distribution allows comparable-mass mergers,
individual merger remnants can reach spins of order
$\chi\sim0.7$. Subsequent coherent gas accretion can further
increase the spin \citep{Volonteri2005,Berti2008,Kritos2024}.}

{If such an accretion episode occurs while the NS companion is already present, the survival of the NS and the retention of a suitable orbit impose additional requirements that could further reduce the occurrence of viable high-spin systems. However, this evolutionary channel is not required by our model, since lower-spin configurations can also reproduce the adopted activity cycle and activity fraction. We therefore do not impose a separate population penalty for rapidly rotating IMBHs. The spin-dependent orbital requirements and system lifetimes are instead understood as part of the uncertainty in $f_{\rm yMag/tight}$.}

{Geometric alignment imposes a strong selection effect. Let $I$ denote the angle between the orbital normal and the line of sight, so that an edge-on configuration corresponds to $I=\pi/2$. For isotropically oriented systems, the probability density is $p(I)=\frac{1}{2}\sin I$ over $0\leq I\leq\pi$. The probability of lying within an angular tolerance $\delta_{\max}$ of an edge-on configuration is therefore}
\begin{equation}
f_{\rm geom}
=\frac{1}{2}\int_{\pi/2-\delta_{\max}}^{\pi/2+\delta_{\max}}
\sin I\,dI
=\sin\delta_{\max}
\simeq\delta_{\max},
\end{equation}
{where the last approximation assumes $\delta_{\max}\ll1$ in radians. Adopting a viewing-angle tolerance of $\delta_{\max}\sim0.05^\circ$--$0.1^\circ$ gives $f_{\rm geom}\sim9\times10^{-4}$--$2\times10^{-3}$. We therefore use $f_{\rm geom}\sim10^{-3}$ as a fiducial orientation factor in Eq.~(\ref{eq:Nobs_final}), keeping its scaling explicit. This viewing-angle tolerance is distinct from the orbital inclination relative to the BH equatorial plane; a full geometrical selection calculation would additionally account for the precession cycle and the required activity-window duration.}
{The most uncertain factor is $f_{\rm yMag/tight}$. This term includes the probability that the IMBH has a bound NS companion, that the NS is young and magnetar-like, that the orbit is compact enough for strong self-lensing, and that the system survives during the active lifetime of the source. 

{We explored two complementary estimates to motivate an
exploratory normalization of $f_{\rm yMag/tight}$. First,
integrating the counts-based ULX luminosity function of
\citet{Swartz2011} above
$L_X=3\times10^{39}\,{\rm erg\,s^{-1}}$ gives a local density
$n_{\rm ULX}\simeq5\times10^{-3}\,{\rm Mpc^{-3}}$.
Extrapolating this density without redshift evolution gives
$N_{\rm ULX}\simeq1.5\times10^8$ within $z=0.5$.
Using the population scaling of \citet{Patruno2005}, with
$\mathcal{R}=200$ and $\tau_X=10^7$ yr, and assuming a viable
magnetar phase of $\tau_{\rm viable}=10^4$ yr, yields
$N_{\rm yMag/tight}\simeq3\times10^7
f_{\rm prog}p_{\rm Mag,tight}$.
Here, $f_{\rm prog}$ selects ULXs associated with relevant
IMBH--NS progenitors, including the adopted mass and
environmental selection, while $p_{\rm Mag,tight}$ selects
descendants that become magnetars and reach the required
orbital configuration while active. For a reference parent
population $N_{\rm IMBH}f_{\rm off/SF}=10^7$, the resulting
fraction is
$f_{\rm yMag/tight}\simeq3f_{\rm prog}p_{\rm Mag,tight}$;
a combined efficiency
$f_{\rm prog}p_{\rm Mag,tight}\sim3\times10^{-4}$
would therefore give $f_{\rm yMag/tight}\sim10^{-3}$.}

{Second, the young-cluster simulations of
\citet{DiCarlo2021} indicate that IMBHs spend approximately
85 per cent of their time in binaries or triples, while NSs
constitute approximately 1.2 per cent of the recorded
companions. If this companion distribution is representative
of the time-weighted occupancy, their product gives an
IMBH--NS occupation fraction of approximately $10^{-2}$.
Assuming that ten per cent of this population consists of
young magnetars in suitable orbits would again give
$f_{\rm yMag/tight}\sim10^{-3}$. This estimate additionally
requires extrapolating from their simulated IMBH masses,
which reach approximately $438\,M_\odot$, to our mass range.
The efficiencies and temporal assumptions used in these
estimates are not established by the cited studies, so the
agreement does not constitute an independent determination
of $f_{\rm yMag/tight}$. We retain $10^{-3}$ as an exploratory
fiducial value with substantial uncertainty and keep its
linear effect on the expected population explicit.
A quantitative determination would require dedicated
population-synthesis calculations.}

With these fiducial values, Eq.~(\ref{eq:Nobs_final}) gives $N_{\rm obs}\sim 1$. {A lower parent IMBH abundance, a smaller geometrical alignment factor, or a formation efficiency of compact IMBH--young-magnetar systems below $10^{-3}$ would reduce this expectation further. Conversely, a higher formation efficiency or a larger fraction of rapidly spinning IMBHs would increase the expected number. Given the substantial uncertainties in these factors, this estimate suggests a rare channel but does not provide a firm prediction of its observable abundance.} A more rigorous estimate would require population-synthesis calculations including IMBH occupation fractions in dwarf galaxies, neutron-star formation and retention, natal kicks, dynamical capture or binary hardening, IMBH spin evolution, and the active lifetime of radio-emitting magnetars{, together with observational selection effects}.

\begin{table}
\centering
\caption{{Dependence on BH spin for
$M_{\rm BH}=10^{3.5}\,M_\odot$,
$M_{\rm NS}=1.4\,M_\odot$, $R_{\rm NS}=10$~km,
$e=0$, $\tau_{\rm FRB}=160$~days,
$\tau_{\rm LT}=320$~days, $\mathcal{A}_T=100$ and an energy fluctuation of 1 order of magnitude of the pulses from the source.
The semi-major axis and inclination are adjusted to
preserve the LT period and an activity fraction of 50\%
according to the approximate analytical window relation.
$\mathcal{A}_{\max}$ denotes the amplification at perfect
alignment and $t_{\rm GW}$ the circular coalescence time.}}
\label{tab:spin_dependence}
\small
\setlength{\tabcolsep}{4pt}
\begin{tabular}{ccccc}
\hline\hline
$\chi$ & $a/r_s$ & $i$ [$10^{-2}$ deg]
& $\mathcal{A}_{\max}/10^4$ & $t_{\rm GW}$ [yr] \\
\hline
0.1 & 192 & 5.27 & 3.66 &  472 \\
0.2 & 242 & 4.69 & 4.11 & 1189 \\
0.3 & 277 & 4.39 & 4.39 & 2042 \\
0.4 & 305 & 4.18 & 4.61 & 2996 \\
0.5 & 328 & 4.03 & 4.78 & 4035 \\
0.6 & 349 & 3.91 & 4.93 & 5145 \\
0.7 & 367 & 3.81 & 5.06 & 6319 \\
0.8 & 384 & 3.72 & 5.17 & 7550 \\
0.9 & 399 & 3.65 & 5.28 & 8834 \\
\hline
\end{tabular}
\end{table}

\end{document}